\documentclass[12pt]{article}

\usepackage[hyphens]{url}
\usepackage[utf8]{inputenc}
\usepackage{charter}
\usepackage{fullpage}
\usepackage{graphicx}
\usepackage{lipsum}
\usepackage[sort&compress,numbers]{natbib}
\usepackage[colorlinks=true,linkcolor=blue,citecolor=blue,urlcolor=blue]{hyperref}
\usepackage[total={6.5in,9in},top=1in,headsep=0.1in,headheight=1in]{geometry}
\usepackage{enumitem}
\usepackage{amssymb}
\usepackage{subfigure}

\newcommand\snowmass{
\begin{center}
  \rule[-0.2in]{\hsize}{0.01in}\\
  \rule{\hsize}{0.01in}\\
  \vskip 0.1in
  Submitted to the Proceedings of the US Community Study\\ 
  on the Future of Particle Physics (Snowmass 2021)\\
  \rule{\hsize}{0.01in}\\
  \rule[+0.2in]{\hsize}{0.01in}\\[-2em]
\end{center}
}

\usepackage[firstpage=true]{background}
\backgroundsetup{contents={\parbox{6.5in}{\snowmass}}, scale=1,placement=top,opacity=1,color=black,position={3.25in,1.2in}}

\usepackage{fancyhdr}
\fancypagestyle{plain}{%
  \fancyhf{}%
  \fancyhead[C]{}
  \fancyfoot[C]{\thepage}
}

\fancypagestyle{empty}{%
  \fancyhf{}%
  \fancyhead[C]{{\it Snowmass 2021 CEF06: Public Policy \& Government Engagement}}
  \fancyfoot[C]{\thepage}
}
\usepackage{authblk}

\usepackage[capitalise,noabbrev]{cleveref} 

\usepackage{comment}
\usepackage{xspace}

\newlength{\myfigurewidth}
\newlength{\myfigureheight}
\newlength{\myfigureskipeqh}
\newcommand{\twofigeqh}[2]{\resizebox{\myfigurewidth}{!}{\includegraphics[height=\myfigureheight]{#1}\hspace{\myfigureskipeqh}\includegraphics[height=\myfigureheight]{#2}}}

\title{Snowmass '21 Community Engagement Frontier 6: Public Policy and Government Engagement\\ Congressional Advocacy for HEP Funding \\(The ``DC Trip'')
}

\date{}

\author[1]{Mateus Carneiro}
\author[2]{Richie Diurba}
\author[3]{Rob Fine}
\author[4]{Mandeep Gill}
\author[5]{Ketino Kaadze}
\author[6]{Harvey Newman}
\author[7]{Kevin Pedro}
\author[8]{Alexx Perloff}
\author[7]{Louise Suter}
\author[9]{Shawn Westerdale}

\affil[1]{Brookhaven National Laboratory}
\affil[2]{Universit\"{a}t Bern}
\affil[3]{Los Alamos National Laboratory}
\affil[4]{Kavli Institute}
\affil[5]{Kansas State University}
\affil[6]{California Institute of Technology}
\affil[7]{Fermi National Accelerator Laboratory}
\affil[8]{University of Colorado Boulder}
\affil[9]{Princeton University}

\begin{document}

\maketitle
\tableofcontents

\section{Introduction}
\label{sec:intro}

This document has been prepared as a Snowmass contributed paper by the Public Policy \& Government Engagement topical group (CEF06) within the Community Engagement Frontier. The charge of CEF06 is to review all aspects of how the High Energy Physics (HEP) community engages with government at all levels and how public policy impacts members of the community and the community at large, and to assess and raise awareness within the community of direct community-driven engagement of the U.S. federal government (\textit{i.e.} advocacy). In the previous Snowmass process these topics were included in a broader ``Community Engagement and Outreach'' group whose work culminated in the recommendations outlined in Ref. \cite{snowmass13recs}.

The focus of this paper is the advocacy undertaken by the High Energy Physics (HEP) community that pertains directly to the funding of the field by the U.S. federal government. The paper is organized as follows. Section \ref{sec:process} describes the mechanism by which the U.S. HEP program is funded within the context of the U.S. federal budget cycle. Section \ref{sec:dc_trip_details} describes the HEP community-driven advocacy effort known colloquially in the community as the annual ``DC Trip''. Section \ref{sec:dc_support} details the supporting community communication materials that are used in these advocacy efforts. Section \ref{sec:future_ideas} outlines specific, actionable ideas for how our community advocacy efforts can be improved on in the near- and mid-term futures. And Section \ref{sec:big_picture_questions} includes a discussion of potential farther-future and larger-scale reforms to science and research funding. 

The authors have written this document to additionally serve as a reference guide for community members about the past and present of the regular HEP advocacy activities in the hopes that the existence of this document itself will be a boon to these efforts. Additional contributed papers prepared by this topical group cover the potential for HEP community advocacy for topics beyond funding~\cite{cef06paper2} and all other aspects of HEP community engagement with government entities \cite{cef06paper3}.

\section{How the U.S. HEP Program is Funded}
\label{sec:process}

To provide some context for HEP community advocacy activities, this section briefly outlines how HEP is funded in the U.S. HEP research is funded primarily through the Department of Energy (DOE) Office of Science (OS) and the National Science Foundation (NSF, and funds are allocated to these agencies in the federal budget on an annual basis. The construction of the budget is a lengthy process that can be roughly broken down into three key steps:

\begin{enumerate}
    \item The President proposes a budget, based on input from executive branch agencies, coordinated by the Office of Management and Budget (OMB).
    \item Congress authorizes and appropriates funding to executive branch agencies considering this proposal and their own priorities. The authorization step dictates what each agency may spend its money on, to variable degrees of specificity, while the appropriations bill ultimately allocates funding.
    \item OMB appropriates funds to the executive branch agencies, as dictated by Congress's authorizations and appropriations bills, to carry out specific programs, projects, and activities in line with the agencies' and the President's priorities.
\end{enumerate}

The President's budget request is formulated using policy guidance from OMB informed by the previous year's budget. Based on this guidance and community input, each agency (DOE and NSF in the case of HEP) submits proposals to OMB. OMB revises and synthesizes these proposals, accounting for Administration priorities, which the President then reviews and transmits to Congress. After OMB has submitted the proposal to the President, there is a period of time during which agencies may appeal revisions made by OMB.

HEP community-led advocacy plays a role during the budget formulation and Congressional steps in this process. In this paper, we cover advocacy focused on the second step enumerated above, \textit{i.e.} advocacy aimed at Congress. We note that input from HEP advisory panels, such as HEPAP and P5, factor into the initial budget proposals sent to Congress and into Congress's deliberations. Current and future community advocacy aimed at OMB and other executive branch agencies is discussed in detail in Ref.~\cite{cef06paper3}. 

\begin{figure}[htb!]
\centering
\includegraphics[width=1\linewidth]{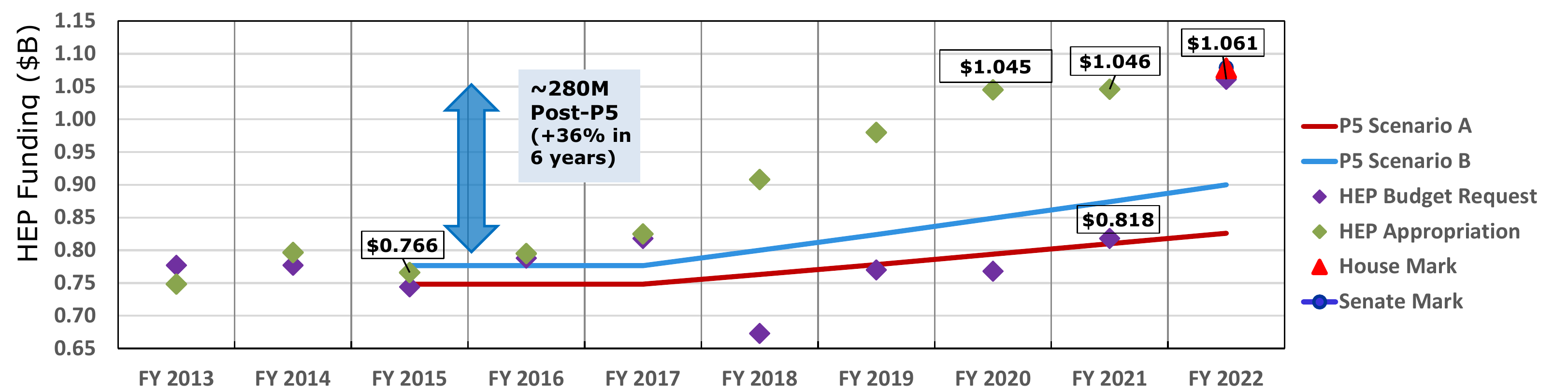}

\caption{DOE OS HEP funding since the last P5, taken from the March 2022 DOE presentation to HEPAP~\cite{march_hepap}. The President's budget request is shown in purple and the Congressional appropriated budget is shown in green. See Table~\ref{tab:request} for the community requests during a similar time frame.}
\label{fig:OHEP_fun}
\end{figure}

The HEP community working with policy experts formulates Congressional appropriations requests for both the DOE OS Office of High Energy Physics (OS HEP) and NSF, which are the two agencies that fund HEP in the U.S. During the appropriations process in Congress, budget narratives provided by the HEP community justify the funding levels in these requests. The community's and President's requests often differ, as does the amount ultimately appropriated by Congress. Figure~\ref{fig:OHEP_fun} shows the President's request, House and Senate proposed funding (prior to reconciliation), and the Congressional appropriations for OS HEP in recent years compared to two of the funding profiles laid out in the 2014 P5 report. Table~\ref{tab:request} compares the President's request, the community request, and Congressional appropriations over a similar time scale. Specific numbers are included for OS HEP only, because HEP is not a dominant fraction of the NSF portfolio.

\begin{table}
    \centering
    \begin{tabular}{c|c|c|c}
    Year & President's Request & Our Ask (\$M) & Appropriated (\$M)\\
    \hline
    FY23 & TBD   & 1356     & TBD   \\
    FY22 & 1,061 & 1180     & 1078  \\
    FY21 & 818   & 1285     & 1046  \\
    FY20 & 768   & 1045     & 1045  \\
    FY19 & 770   & N/A$^{*}$& 980   \\
    FY18 & 673   & 860      & 908   \\
    FY17 & 818   & 833      & 825   \\
    \end{tabular}
    \caption{The community request and Congressional appropriations in millions of dollars. *Due to delays in the FY19 budget cycle, no advocacy aligned with this budget.}
    \label{tab:request}
\end{table}

\section{Details of Current HEP Community Advocacy}
\label{sec:dc_trip_details}

Every year a group of physicists and other members of the HEP community travels to Washington, D.C. to share excitement about their research and to foster support for HEP with policy makers. This annual advocacy effort aims to visit with as many members of Congress (and their staff) as possible and with Administration and funding agency representatives. Coordination of the trip is a joint effort by Fermilab Users Executive Committee (UEC), SLAC Users Organization (SLUO), and the U.S.-LHC Users Association (USLUA). Input to this effort is additionally solicited from the Executive Committee of the American Physical Society Division of Particles and Fields (APS DPF).

The goal of this effort is to garner support for the funding of physical sciences research in general and for HEP in particular. Members of the delegation discuss the overall benefits of HEP and basic research with policy makers and deliver a specific appropriations ``Ask'' for HEP funding. This ``Ask'' includes a request for specific funding levels for each of DOE OS HEP and for NSF. The timing of this trip is usually chosen to align with moment in the annual budgetary cycle when the budget is being discussed by Congress (typically this is in March).

\subsection{``DC Trip'' Logistics: Planning}

The ``DC Trip'' has existed in some form for more than 35 years, and has evolved over that time into the current effort as described in this document. Historically, facility users at Brookhaven National Laboratory (BNL), the Stanford Linear Accelerator Collider (SLAC), and Fermilab (FNAL) executed policy advocacy efforts independently of one another on an as-needed basis, not as a coordinated annual activity. FNAL and SLAC were the first to coordinate, seeding the effort that exists today, though in recent years SLAC participation has diminished relative to strong FNAL participation. The UEC represents nearly 4,000 FNAL users, esentially all of whom are supported in their research through HEP funding. SLAC has around 3,000 facility users, but the majority of those are not currently supported by HEP funding. Since its formation (in the early 2000's) USLUA has actively participated in this effort. In 2013 USLUA signed an agreement with the Universities Research Association (URA) and was registered by DC.gov as a nonprofit association in Washington, D.C. at URA headquarters. USLUA has over 1,000 members. Specific data about the effort, including participation metrics, only exists as far back as 2010 -- these data are preserved and documented on a dedicated wiki, maintained by the UEC.

The chair of the subcommitte on Government Relations in the Fermilab UEC is responsible for leading the trip each year and coordinates most of the trip's logistics. During the period since the last Snowmass in 2013, the following community members served in this role: Nadja Strobbe (2022), Ketino Kaadze (2021), Kirsty Duffy (2020), Fernanda Psihas (2019), Joseph Zennamo (2018), Louise Suter (2017), Jesus Ordu\~{n}a (2016), Sandra Biedron (2015), Breese Quinn (2014).

URA is a leading contributor to funding for the annual advocacy effort. Each year the UEC submits a funding request to URA for support, which has historically provided support for about 70\% of trip participants. This funding is first allocated to UEC-invited members, and then as much as possible to support additional participants who don't have other available funding. Stanford provides funds for the SLUO group. The majority of USLUA members pay out-of-pocket. The small funding available at USLUA supports early career community members invited by USLUA.

The members of UEC, and the executive committees of USLUA and SLAC make up the core delegation that travels to Washington, D.C. for this effort. These individuals, who are all chosen through election, represent a large fraction of the U.S. HEP user community. A number of additional community members invited by these groups rounds out the delegation each year. In practice, recruitment of trip attendees is performed using various tactics. For example, USLUA invites early career awardees from its annual U.S. LHC meeting, and UEC invites the elected representatives of the Fermilab Student and Postdoc Association. The trip attendees are distributed as 50\%, 35\%, and 15\%, between UEC, USLUA, and SLUO, respectively. Attempts are made to recruit people from different career stages, demographics, and research areas. Approximately one-third of participants are postdocs, one-third are professors or national laboratory scientists, and the final third is distributed between Ph.D. students and other university or laboratory staff~\footnote{In recent years, a survey has been distributed to attendees of the trip after the fact -- these figures are based on the demographic information available in responses to these surveys in the last few years.}. The majority of participants represent the Energy Frontier and Intensity Frontier (as defined by the 2013 Snowmass process). While efforts have also been made to have broader representation in the trip from the Theory, Cosmic, Computational, Instrumentation communities, there is room for improvement in this aspect of the organization.

\subsection{``DC Trip'' Logistics: On the Ground}

Until the COVID-19 pandemic caused the community to transfer to a virtual ``trip'', this effort had been an in-person trip to Washington, D.C. When occurring in-person, the trip takes place over three days, during which community members visit the offices of Members of Congress, generally meeting with Congressional staff, although meetings with AAAS fellows and Representatives or Senators are not uncommon. The meetings are generally attended by pairs of community members, a ``Primary'' who has some connection to that office and responsibility for coordinating the meeting, and a ``Secondary'' that helps to facilitate a constructive dialogue during the meeting. A day full of these meetings is both physically and mentally exhausting -- often trip attendees spend all day in meetings, interspersed with typically miles of walking between offices on Capitol Hill. Meetings range from 5 minutes to an hour but generally are around 20-30 minutes long.

Trip attendees are provided with a packet of materials to bring to each of the meetings for which they are the ``Primary''. The packets contain various leaflets and brochures about HEP, how supporting HEP benefits society, and other materials that can help to facilitate conversation during these meetings. Arguably the most important element of the packet is ``the Ask", a document which very clearly and succinctly spells out the level of funding that the community is asking Congress to appropriate to DOE OHEP and for NSF, though in the later case no specific allocation can be made for HEP. In addition, various small keepsakes have been brought in the past including rulers made of scintillating plastic, ``Particle Zoo'' buttons, and cardboard augmented reality viewers that link to virtual tours of Fermilab. A more detailed discussion of these materials (and ideas for how to improve them) is presented in Section \ref{sec:dc_support}.

The flow of a conversation during one of the meetings between trip attendees and Congressional staff depends largely on the personal experiences and interests of those in the meeting, and there can be a large variance in how meetings proceed between different offices in the same year and between the same office in different years. The specific topics brought up by trip attendees for conversation are often driven by the perceived priorities of the Member of Congress and their office. Additionally the political climate and priorities of the Presidential administration can effect the delivery of the high-level message each year. There are a number of common arguments that are made during these meetings while presenting the case for strong and sustained HEP support. Examples of such arguments follow.

\begin{itemize}
    \item \textbf{STEM education} and maintaining a strong STEM workforce in the U.S. are issues that resonate strongly with government representatives. Details about the community's efforts on various outreach activities, attracting new generations to pursue research in HEP, and the success of HEP graduates both within and outside of the field, in other research activities, or in industry, are generally well received.
    \item \textbf{Attracting new talent to the U.S.} from around the world is generally perceived in Congress to be a desirable outcome of maintaining a strong science program. Greater available funding means attracting and retaining more highly talented individuals, whom are generally better prepared to tackle future national and international challenges in science, technology, and security.
    \item The notion that \textbf{basic research drives technological developments} is a core tenet of our advocacy. While advancements that follow from investments into basic research are generally not apparent immediately, we have significant evidence to indicate that these investments reliably have a profound impact on society in the long term (e.g. through quality of life advancements).
    \item \textbf{District and state-level impact} is discussed in the context of specific information on HEP grants and HEP procurement spending. Each year, as part of the organization of the trip, databases that track these to the level of Congressional district (through zip codes) are updated. These data can be very impactful for providing concrete examples of how HEP funding has both broad impacts across the country and local impact in specific districts. 
\end{itemize}

In addition to the Congressional office meetings, meetings are also organized with the staffs of the (sub)committees that most impact HEP funding and policy. These are the House and Senate Commerce, Justice, Science, and Related Agencies Appropriations Subcommittees, the House and Senate Energy Appropriations Subcommittees, the House Research and Technology Appropriations Subcommittee, the Senate Space and Science Appropriations Subcommittee, the House Energy and Water Development, and Related Agencies Appropriations Subcommittee, and the Senate Energy and Water Development Appropriations Subcommittee.

\subsection{The Washington-HEP Integrated Planning System}

The Washington-HEP Integrated Planning System (WHIPS) is a framework developed to handle most of the logistics of planning, executing, and documenting the ``DC Trip''. Aspects of the trip planning logistics were developed by many individuals over many years, and a considerable effort was undertaken by Fernanda Psihas and Justin Vasel in 2019 to consolidate these aspects into a unified framework. WHIPS has significantly improved the efficiency of organizing, and thereby the success of, the annual HEP community advocacy effort. The framework tracks:

\begin{itemize}
    \item trip attendees and their connections to Congressional districts/offices,
    \item members of Congress and their committee assignments,
    \item Congressional office meetings,
    \item Congressional committee meetings,
    \item executive branch meetings,
    \item post-meeting reports,
    \item Fermilab procurement data,
    \item DOE and NSF grant awards data,
    \item DOE SULI (Science Undergraduate Laboratory Internships) and CCI  (Community College Internships) student internship programs data, and
    \item historical trip performance metrics.
\end{itemize}

Connections between trip attendees and Congressional districts and offices form the basis for determining the ``Primary'' assignments each year. These connections represent some form of relationship between a community member and a Member of Congress. For example, each trip attendee has a connection by virtue of the Congressional district and state that they live or work in. Therefore, each attendee generally has \textit{at least} three connections (two Senators and one Representative). There are ten fundamental types of connections that WHIPS tracks as of the 2022 advocacy effort:
 
\begin{enumerate}
    \item Registered Voter,
    \item Current Resident,
    \item Former Resident,
    \item Immediate Family,
    \item Extended Family,
    \item Current Workplace,
    \item Former Workplace,
    \item Educated in District/State,
    \item Personal Connection, and
    \item None of the Above.
\end{enumerate}

It is common for several of these types to apply to a single connection. Each type has an associated score, and the combination of these scores is used to quantify the overall connection strength that a particular trip attendee has to a particular office. The composite connection score is essential to optimizing the assignment of meetings to the attendees of the trip each year.

Each trip attendee has a profile in which they enter their biographical information, which is useful for other trip attendees they haven't met, and all of their connections to Congressional offices. Once all of the profiles have been created, the trip organizers and WHIPS administrators run an algorithm to score the connections and make meeting primary assignments accordingly. It is then up to the attendees to schedule meetings and to enter the relevant meeting information, scheduling notes, and post-meeting notes into WHIPS.

Another feature provided by WHIPS is the automated creation of district- and state-specific information. Before each trip, the WHIPS developers enter information about each of the below items for each Congressional district.

\begin{itemize}
    \item Fermilab procurements made during the preceding fiscal year, broken down by state and district. This information is provided annually by Fermilab.
    \item DOE and NSF grant awards. This is publicly available information.
    \item Participants in DOE's SULI and CCI internship programs. This is publicly available information. 
\end{itemize}

This information is scraped from a variety of sources, and is aggregated using the HEP funding webpage/grants database created by Michael Baumer \cite{baumer}.\footnote{We note that these types of data-scraping tools and databases have a long history of use in HEP community advocacy efforts. This particular tool is the latest iteration and is the one that is currently in use.} This information is then entered into WHIPS, which produces documents specific to each district or state for use during meetings. Anecdotal feedback from Congressional staff indicate that these documents about district/state-level funding and student participation levels are very helpful for connecting basic science funding to specific impacts at the district and state levels.

WHIPS has had a substantial and positive impact on the planning and organization of the annual HEP advocacy trip. This framework will likely continue to be used for years, if not decades to come. \textbf{We note the need for work on the management model for using and maintaining WHIPS.} The framework is currently maintained by its original developers, Fernanda Psihas and Justin Vasel. However, their participation in community advocacy efforts and availability are not guaranteed indefinitely. Therefore, it is prudent to expand the number of people who can maintain the framework and continue its development. Ideally, these new people would have fairly stable positions within the HEP community and therefore be more likely to remain available for years to come. 

\subsection{The ``DC Trip'' Wiki}

The UEC, SLUO, and USLUA also collectively maintain a wiki \cite{DC_trip_wiki} that has information on the trip going back to 2010. This resource serves as an information repository and historical reference. The contents of the wiki generally grows in breadth and depth each year. Below is a non-exhaustive list of the wiki's current contents.

\begin{itemize}
    \item General information about the DC trip.
    \item Training materials useful for first-time participants and for those that need a refresher (see section~\ref{sec:dc_trip_training}).
    \item Information which is helpful for scheduling meetings.
    \item Easy access to the packet materials.
    \item Information about interesting physics programs and connections that might be mentioned during the meetings.
    \item A historical repository of the records from past trips.
\end{itemize}

Right now, the wiki is not a public resource; only trip attendees have access. Much of the information on the wiki is publicly available elsewhere and non-controversial in nature. However, other items, such as the ``Ask'', is kept private until the beginning of the trip because it is liable to change up until the last minute and it's important that only a single version of it be available to ensure unity of messaging. The idea of transitioning this resource to a public website with a more nuanced protection scheme has been discussed. It is worth considering who would host and pay for such a new website. Presently the page is sponsored by URA and managed by the Fermilab UEC. Depending upon the management of the DC trip in the future, it may be appropriate to revisit the management and maintenance of the wiki.

\subsection{Training for the ``DC Trip''}
\label{sec:dc_trip_training}

The ``DC Trip'' in its present state is a significant endeavor involving $\mathcal{O}(100)$ people from across the United States. These volunteers have varying levels of understanding of the U.S. budget and appropriations process, how to communicate about HEP, how to interact with policy makers, and the day-to-day and meeting-specific logistics of the trip. As such, it is necessary that training on these topics be provided to trip attendees in a short period of time.

There are typically two or three dedicated training sessions in advance of the trip in a given year. These sessions begin approximately one to two months before the start of the trip. Many topics are covered, including but not limited to an overview of the U.S. budget process, how to effectively communicate about science, the specifics of the ``Ask'' of that year, the use of WHIPS, how to schedule meetings, and what to say during meetings. This is a large amount of information to impart over a short period.

To supplement the lecture-style training sessions, there are some example videos demonstrating how to approach meetings. These have been produced by experienced participants of the trip and in aggregate cover a variety of styles and scenarios reflective of the spread in how these meetings can take place. As of the 2022 effort, this repository has been expanded to include videos demonstrating how to conduct remote meetings (\textit{i.e.} using Zoom or another teleconferencing system), which has notably different dynamics compared to in-person meetings. While these videos will eventually become obsolete, the idea is that they are more or less independent of the specific year in which they were produced and will serve as a useful resource for years to come.

The set of topics covered by the training sessions and video tutorials is inherently incomplete, and trip organizers have long recognized the importance for first-time participants to get ``hands-on'' experience. Therefore, the recommendation is generally that individuals new to the trip attend at least one, if not more, of their early meetings as a ``secondary'' participant rather than as the primary contact for that meeting. This allows the less experienced attendee to gain some first-hand experience while not saddling them with too much responsibility right from the start. This encouragement is provided early and often throughout training and leading up to the trip each year, but it is ultimately the responsibility of each trip attendee to schedule meetings with the offices for which they have been assigned as ``Primary''.

\subsection{Metrics of the ``DC Trip'' Impact}
\label{sec:metrics}

The percentage of offices visited has grown over the years as financial support for the trip, and therefore attendance on the trip, has grown. In 2010 there were 34 attendees; by 2017 this number grew to 50. In 2017 about 70\% of House and Senate offices were visited. In 2019 URA started to provide additional funding for the trip, meaning that up to 70 attendees could be supported, and in that year 100\% of House and Senate offices were visited. This was the most recent in-person trip as of the writing of this document, due to COVID, and during recent years the reach (by percentage of offices ``visited'') has been significantly lower.

It is hard to quantify the direct impact of our advocacy effort, because the main aim of the annual trip is to increase awareness of and excitement about HEP within Congress. Strong anecdotal evidence indicates that we are having success in this respect -- repeat attendees have built long-lasting connections to offices and staffers often remember past visits and have said that they enjoy meeting with us and getting the opportunity to ask about HEP. It is similarly difficult to judge our success on affecting the federal budget. Many factors affect the overall funding profile for HEP, based on the priorities of the Administration and Congress, and influenced by current affairs. Therefore it is hard to map the appropriated funds for DOE OS and NSF directly back to the impact of our advocacy. Figure \ref{fig:OHEP_fun} shows that the HEP funding profile has grown since the last P5, but many factors go into that increase. Again, anecdotal evidence -- positive feedback from the funding agencies and from Congress -- indicates that community advocacy effort \textit{has} contributed to the current funding levels for the field.

One of the more quantitative metrics for the success of the DC trip is the number of signatories on ``Dear Colleague'' letters. These letters are drafted by Senators or Representatives, directed to their peers, and request that they sign to show their support for some particular legislation or position described in the letter. During the appropriations process, there are many such letters in support of various specific funding priorities, often with particular requested funding values, that are sent to the appropriations (sub)committees.

Each year there are a pair of these letters in each of the House and Senate asking for support of DOE OS and for NSF. The details of these letters are a common point of conversation during meetings and attendees are usually instructed to ask  members of Congress to consider signing onto them. The numbers of signatories on these letters for the last three years are given in Table~\ref{tab:dcl}. Approximately 30\% of the Senate signs the DOE OS letter, and 40\% the NSF. Approximately 25-35\% of the House signs the DOE OS, and 35-40\% of the House signs the NSF letter. We discuss ``Dear Colleague'' letters in a broader context in Section~\ref{sec:other_comm_hep}.

\begin{table}
    \centering
    \begin{tabular}{c|c|c|c}
    & FY22 & FY21 & FY20 \\
    \hline
    DOE House   &   112 &   142 &   160 \\
    NSF House   &   155 &   177 &   173 \\
    DOE Senate  &   30  &   31  &   30  \\
    NSF Senate  &   41  &   40  &   37  \\
    \end{tabular}
    \caption{Number of signatories on the ``Dear Colleague'' letters in support of  DOE OS and NSF in recent years.}
    \label{tab:dcl}
\end{table}

The question of identifying additional metrics that could be tracked to gauge the impact and success of the ``DC Trip'' came up repeatedly throughout the proceedings of CEF06. No specific metrics were identified. More investigation and discussion in this area would be very beneficial. 

\subsection{Community inreach}

The planning and execution of the ``DC Trip'' are generally contained within a small group ($\sim$100) of the members of the users organizations (UEC, SLUO, and USLUA), DPF, and past trip attendees. It has been discussed that having more knowledge and discussion of the activities in various formats would benefit the field at large and have a positive impact on the community's advocacy goals. For example, broader inreach into the community would help in to have a more representative group attend the trip and would likely provide stronger connections to every district.

The current state of community inreach about this advocacy is very limited. There are a limited number of updates provided to the community each year at APS DPF and user group (Fermilab, US-HEP) annual meetings, as well as annual updates provided to HEPAP (beginning in 2017, see Refs \cite{hepap_2017, hepap_2018, hepap_2019, hepap_2020, hepap_2021}) at public meetings. Additionally some amount of details can be found through User group meeting minutes, reports, and websites. Overall, the reach of these means of communication into the community is limited, and awareness of advocacy efforts is generally very low. 

\section{HEP Community Advocacy Materials}
\label{sec:dc_support}

A key element of meetings during the ``DC Trip'' is a series of informational pamphlets designed to visually relay the community's most important messages. These materials serve a dual purpose, which is to both provide a conversation piece during the meetings and to leave a reference guide for the Congressional staff as they make their recommendations.

The following subsections describe how these documents are produced and by whom, what documents are currently available, and discuss additional resources which may be beneficial for discussions between community members and Congressional staff and other policy makers. 

\subsection{HEP Community Communication Materials}
\label{sec:comm_materials}

There is an annual process to update a centrally hosted set of general-purpose (and publicly available) HEP community materials. These materials can be found at the US particle physics website \cite{uspp2022mar}, and are jointly maintained by APS DPF, UEC, SLUO, and USLUA. These community communication materials are an essential part of the Congressional HEP advocacy strategy but are separate from it in many key ways. They serve the larger purpose of communicating about HEP and its benefits, and they are used in multiple forums of which the advocacy efforts are just one. We note that due to the Hatch Act of 1939 government employees are expressly barred from engaging in political activities, so these materials have always been explicitly created as communication material and explicitly not to advocate for any area of government support. Their purpose is to inform people in general about HEP and its benefits. 

While these materials serve a broader purpose, the annual ``DC Trip'' generally provides the impetus for them to be updated on a regular basis. The process of updating existing documents and/or creating new documents begins approximately five to six months prior to the start of the ``DC Trip'' each year. A committee of volunteers is formed by representatives from UEC, SLUO, and USLUA, with support from the Fermilab Office of Communication, DOE, and the former P5 chair. Each year this group is reformed with a somewhat new membership, which is a challenge when it comes to maintaining best practices and staying consistent in messaging year-over-year. In the recent past, the community has received support from a member of DOE who has helped the community coordinate the drafting of these materials. This individual acts both as a repository of knowledge and as the group leader, making sure that steady progress is made on schedule. During the 2021-2022 session, guidance was also provided by a member of the Science Communication team within DOE OHEP. This individual has been helpful in crafting coherent, impactful, and professional messaging throughout the documents.

We note that the contributions of various individuals have been extremely helpful in crafting the trip materials, but that there is no formal requirement that this support be provided from the Fermilab Office of Communication, DOE, and the former P5 chair. There is a risk that this support may someday end and it would behoove the community to find some other way of maintaining records and maintaining support from these organizations and individuals.

The role of the Fermilab Office of Communication is to help in crafting a professional-sounding and professional-looking message, as well as putting the group in contact with layout and design professionals who can design high quality documents. Typically the layout professional is contracted at the start of the process, but is not directly involved until the end of the process, when all of the textual updates are completed.

In 2016 there was large push to get all of the main community materials updated to have consistent and professional formatting. Since then, during any given year only a few documents need to be updated and up to one or two new documents are created. One of the first jobs of the committee is to decide which (if any) of the documents needs to be updated, if there is a need for a new document, or if any should be retired. Typically, the updates are fairly straightforward, while the creation of a new document takes much more planning and community input.

One of the key documents, which is updated every year, highlights the HEP community's progress (in the context of the P5 plan) during the past calendar year and its priorities for the coming year. The group works collaboratively to decide how best to relay the community's top priorities and to remain consistent with the messaging of the P5 report. This process is typically led by the previous chair of the P5 committee due to their ability to balance the long-term strategic plan with the recent and current progress and priorities.

Sometimes trip attendees or community members see the need for a new document to explain a given experiment or field of research. In these cases, it's common for experts in the given field to help in the creation of the new document. Once the document is created, it is the responsibility of future committees to review the document for needed updates or to solicit feedback from experts on the subject matter who can provide guidance on any updates that need to be made.

The needed changes to the text are eventually collected, at which point these drafts are sent to experiment, laboratory, and group leaders as well as other notable individuals (and groups) within the HEP community. Once feedback is received from these individuals (and groups), a final set of drafts is sent to the layout professional, at which point there are several iterations of proofs and comments. Once the \textit{printed} layout is finalized, a digital version of the information is also created, which requires its own review process.

Below is a list of the various documents produced and the aspect of HEP that each seeks to explain.

\begin{itemize}
    \item \textbf{Particle Physics Progress and Priorities.} Regularly updated, this document examines the HEP community's recent accomplishments and the priorities for the upcoming year. 
    \item \textbf{Particle Physics is Discovery Science.} This document provides basic information about participle physics, listing the broad questions that the HEP community is trying to answer and how individuals go about doing that, with ties back to big questions called out in the last P5 report. 
    \item \textbf{Particle Physics Makes a Difference in Your Life.} The discoveries and technological developments made by particle physicists have an impact on the everyday lives of ordinary citizens. We know that this type of broad applicability can resonate strongly with members of Congress and their staff. Therefore, this document discusses some ways in which particle physics has impacted other fields and industries.
    \item \textbf{Particle Physics Builds STEM Leaders.} This document discusses the outreach and public engagement activities of the field. While education is no longer a ``top-ten'' issue for voters \cite{pew2020aug,gallup2022mar}, it is a consistently important topic among their elected representatives and federal funding agencies.
    \item \textbf{Particle Physicists Value Diversity and Strive Toward Equity.} Most importantly this document contains a statement of HEP community values and describes current goals in terms of diversity, equity, and inclusion. The document very clearly states that the field has work to do towards improving the climate and opportunities for historically underrepresented groups (HUGs). The document also highlights a few programs which are designed specifically to empower and provide opportunities for members of HUGs.
    \item \textbf{Particle Physicists Deliver Discovery Science Through Collaboration.} This document discusses the various institution types and locations where particle physicists work and conduct research. The goal is to show that HEP is an international effort and involves some of the best minds from across the country and world. This document also contains a summary of the projects in the 2013 P5 report, their timelines, and whether or not they are on schedule.
    \item \textbf{Particle Physicists Advance Artificial Intelligence.} Although the HEP community has been working on machine learning (ML) for decades, artificial intelligence (AI) only became a national initiative in March of 2020 \cite{naiia2020,aigov2022}. This document was created to explain how the HEP community uses ML and successfully interfaces with partners in industry to push the boundaries of ML research and development.
    \item \textbf{Particle Physics and Quantum Information Science.} Quantum Information Science (QIS) is another area of national interest \cite{osti_qis_2022mar}. This document discusses the benefits of QIS, the skill sets that make particle physicists valuable in these endeavors, and how QIS developments can, in turn, help solve fundamental problems in HEP.
    \item \textbf{Particle Physics in the U.S. Map.} This map shows the distribution of institutions involved in HEP and that receive funding from either DOE or NSF. While absolutely valued as members of the HEP community, this map does not contain institutions who receive funding from non-federal or non-HEP sources. This is solely because of the challenge in finding information about these types of interdisciplinary grants.
\end{itemize}

All of these documents are hosted (and available for download) at Ref.~\cite{uspp2022mar}. In addition, this website provides online versions of each of these documents. The contents of the website are updated each year at the same time as the various documents. This process generally concludes in the Spring of each year, so that the materials have been fully updated in advance of the annual advocacy trip.

\subsection{``DC Trip'' Advocacy ``Packet'' Materials (2021 Trip)}

In addition to the community-wide information described in Section~\ref{sec:comm_materials}, a series of informational pamphlets about specific fields of study, national laboratories, and other programs are also utilized as part of ``DC Trip'' meetings each year. While useful for highlighting specific institutions or areas of research, these additional  documents are typically very short, only about one page in length, and are generally not the primary reference documents during any individual meeting due to their high level of specificity. These additional materials are summarized below.\\

\noindent Materials detailing the facilities of the three users groups organizing the trip:
\begin{itemize}
    \item \textbf{Fermi National Accelerator Laboratory.} This pamphlet is included to show the breadth of research taking place at the only national laboratory exclusively dedicated to particle physics.
    \item \textbf{The United States at the Large Hadron Collider.} A document specifically designed to talk about the upsides to supporting the Large Hadron Collider (LHC) projects.
    \item \textbf{SLAC National Accelerator Laboratory.} Like the Fermilab pamphlet, this document covers the HEP related science taking place at SLAC.
\end{itemize}

\noindent Materials summarizing science drivers from P5:
\begin{itemize}
    \item \textbf{Cosmic Acceleration.} This pamphlet contains a discussion of dark energy and how the HEP community plans to study its mysterious effects.
    \item \textbf{Dark Matter Science: Beyond the Ordinary.} A brief overview of dark matter and a non-exhaustive list of experiments which seek to study it.
    \item \textbf{Understanding Nature's Most Mysterious Particle.} This document discusses the neutrino and the open mysteries surrounding the particle.
\end{itemize}

\noindent Project or Program specific:
\begin{itemize}
    \item \textbf{Building an International Flagship Neutrino Experiment.} This document discusses the Deep Underground Neutrino Experiment (DUNE) and associated projects, the Long-Baseline Neutrino Facility (LBNF) and the Proton Improvement Plan II (PIP-II).
    \item \textbf{Vera C. Rubin Observatory: The Legacy Survey of Space and Time.} This document briefly explains the science goals and construction of the Rubin Observatory's Legacy Survey of Space and Time (LSST).
    \item \textbf{VetTech Program at Fermilab.} This pamphlet was created prior to the passage of the National Defense Authorization Act of 2020~\cite{ndaa2020} and the Department of Energy National Labs Jobs ACCESS Act~\cite{nljaa2020}. It highlights the hugely successful program at Fermilab, which was the model used for the creation of the nation-wide program.
\end{itemize}

Fermilab (and hosted projects) specific material is available at Ref.~\cite{fnal_factsheets}, and
SLAC (and hosted projects) specific materials is available at Ref.~\cite{slac_factsheets}.

The final document utilized in advocacy meetings, though arguably the most important, is ``the Ask". As noted above (in the discussion of ``DC Trip'' logistics) this document very clearly and succinctly spells out the level of funding that the community is asking Congress to appropriate to DOE OHEP and for NSF, though in the later case no specific allocation can be made for HEP. The specific levels of funding requested in ``the Ask'' are determined in consultation with government relations experts (currently Lewis \& Burke Associates), and all HEP stakeholders. 

\section{Other HEP Engagement of Government Officials} 
\label{sec:other_comm}

In addition to the annual advocacy trip and related activities, members of the HEP community engage in a variety of other interactions with Congress. Some of these additional activities occur more regularly than others, and some include written communications while others are exclusively in-person interactions. And some of these activities are specifically coordinated by the HEP community or otherwise involve individuals explicitly representing the HEP community, while others are activities coordinated by non-HEP groups that members of the HEP community have participated in. These activities, categorized by this latter characteristic, are described briefly below. Connections to the activities of the annual advocacy trip are noted where relevant.

\subsection{HEP-Community-Specific Activities}
\label{sec:other_comm_hep}

\paragraph{HEP community appropriations requests.} The HEP community regularly sends letters to the leading members of the House and Senate appropriations (sub)committees in charge of HEP funding. Specifically these are the Energy and Water Development (E\&W) Subcommittee, which is in charge of appropriations for DOE, and the Commerce, Justice, Science, and Related Agencies (CJS) Subcommittee, which is in charge of appropriations for NSF. These letters are written anytime there is a perceived difference in the budget compared to community expectation and are tailored to handle the specific funding landscape.

\paragraph{``Dear Colleague'' letters.} It is routine for ``Dear Colleague'' letters to be written by members of Congress who support a particular piece of legislation. These letters explicitly voice that support and then are signed by other Members who also support that legislation. The letters themselves, when the relevant legislation is budgetary, are addressed to the leading and ranking\footnote{The ``ranking'' member of a congressional (sub)committee is the leader of the minority party.} members of a particular appropriations subcommittee. While not originating from within the HEP community, we take it upon ourselves to encourage as many Members to sign onto these letters as it shows strong and broad support for the funding levels we are requesting. In any given year there will be four letters that are of note to the HEP community, two each for the House and the Senate, each with one letter for the E\&W Subcommittee and one for the CJS Subcommittee in each body. As noted in Section \ref{sec:metrics}, one of the regular activities of the ``DC Trip'' is to encourage Members to sign onto the letters.

\paragraph{Programmatic (appropriation) requests.} As part of the process of collecting input about a particular funding bill, Congressional offices routinely collect what are called ``programmatic requests'' from their constituents. These requests call for specific levels of funding for a given appropriations bill. While there is no standard way of collecting this information (\textit{i.e.} Google Form, website, email, etc.), they are all collected roughly around the same time as the ``DC Trip'' (\textit{i.e.} usually in late March). As part of organizing the trip, attendees are instructed to ask their assigned offices how they like to collect these programmatic requests and to submit requests as one of their core responsibilities. The text and funding levels specified in these requests are the same as in the letters sent to the appropriations committees to ensure unity of messaging.

\paragraph{Written Congressional testimony.} In the past, the HEP community has submitted written Congressional testimony in support of particular appropriations levels. However, this is a practice which has fallen out of favor and has been discouraged by the government relations experts advising HEP community advocacy efforts. Writing Congressional testimonies and submitting them to the appropriations subcommittees is a lot of work, especially to keep it consistent within all of the other constraints and pieces of communication.

\paragraph{Verbal Congressional testimony.} Verbal congressional testimony is another tool which has not been used in many years. Verbal testimony is by invitation only, with speaker selection done by the committee itself. This raises question on controls available to the community to make sure that the individual invited stays ``on message'' and that their testimony is consistent with other Congressional communications from the community. It has been, and continues to be, vital that the entire community support the P5 plan with uniform messaging. In the past Congress has conveyed its unhappiness when they hear conflicting messages from the HEP community. The last time that someone from the HEP community was invited to provide verbal testimony was on 3 May 2017 when Dr. Maria Spiropulu (Caltech) spoke to the House Energy \& Water Appropriations Subcommittee. Dr. Spiropulu was, at the time, a member of HEPAP. Among the areas that she highlighted were the neutrino physics program in the U.S., including LBNF/DUNE; the dark energy and dark matter experimental programs; and U.S. support for the LHC. She also described the contributions of HEP to technical innovations and a well-trained scientific workforce.

\paragraph{Higgs Day on the Hill.} The HEP community organized the ``Higgs Day on the Hill'' \cite{higgs}, which was a one-time activity following the discovery of the Higgs boson. Other professional societies and organizations also host activities which are somewhat related to the needs and activities of the HEP community and that sometimes include members of the HEP community.

\subsection{Non HEP-Community-Specific Activities}

HEP community members have participated in various other efforts to engage government officials over the years. A subset of these is described briefly below.

\begin{itemize}
    \item Prior to COVID-19, the government relations team for the Society for Science at User Research Facilities (SSURF) held their own yearly advocacy trips. Each of the national laboratories and research facilities was nominally represented on this trip, accompanied by government engagement professionals. This trip included a reception for staffers on the Hill and at one point included a demonstration using an Oculus Rift to show event displays from MicroBooNE (a neutrino experiment that operated at Fermilab from 2015-2021).
    \item Representatives of Congressional offices (generally staff, and sometimes Members) on appropriations subcommittees occasionally visit experimental sites to see how funds appropriated funding agencies is utilized. During such visits researchers are generally asked to engage with these visitors, which provides an excellent opportunity for direct constructive interaction between researchers and legislators.
    \item The American Astronomical Society (AAS) holds its own Congressional Visits Day (CVD)~\cite{aas_cvd_2022mar}.
    \item APS and APS DPF regularly hold ``write to your Congressperson'' drives in which a website is set up to make it easy for concerned citizens to send a letter to their Congressperson.
    \item Other activities organized by the APS government affairs team~\cite{aps_gov_affairs_2022mar}. 
\end{itemize}

\section{Building on Existing HEP Community Advocacy}
\label{sec:future_ideas}

Throughout the course of conversations organized by CEF06, a number of ideas have been put forth for how HEP community advocacy of the budget process can be improved upon. Additionally, community members that have participated in these advocacy efforts in recent years have made suggestions for improvements to the process independently of Snowmass proceedings. This section documents the subset of these ideas that have been suggested which the authors have determined are inherently near-term or otherwise actionable in nature. A non-intersecting subset of suggestions, which call for longer-term actions or more fundamental changes to the model of HEP community-driven advocacy, are described in Section \ref{sec:big_picture_questions}.

\subsection{Continuity and Succession of ``DC Trip'' Leadership}

The leadership of the ``DC Trip'' has historically been provided by the UEC, SLUO, and USLUA, directed by the chair of the UEC subcommittee on government engagement. This model has its strengths, namely enabling the rotation of the time-intensive coordination of logistical aspects of the effort among community members annually while simultaneously providing the basis for institutional knowledge to be preserved within the three users associations. In practice, however, there are many nuances to the planning of the annual advocacy effort that rely on the expertise of community members who have participated in this effort repeatedly over many years and have developed considerable experience. Examples of this nuance include sustained long-term relationships with Congressional subcommittee staff and experience communicating about legislative details with career staff members at OMB and OSTP.

It has been noted repeatedly by community members that the model in use is susceptible to single points of failure in the form of individuals relied upon for a disproportionate amount of subject matter expertise. Specifically, it has been suggested that a plan be developed to intentionally expand the pool of community members that have specialized expertise in the following key advocacy areas.

\begin{itemize}
        \item Organizing meetings with funding agencies and executive branch agencies during the annual trip to Washington, D.C.
        \item Organizing and conducting meetings with (sub)committee staff during the annual trip to Washington, D.C.
        \item Development and administration of WHIPS and other technological elements of the advocacy infrastructure.
\end{itemize}

It is generally agreed that the architecture of the present scheme is sufficient for coordinating general logistics for the ``DC Trip'' (though we note that this does not preclude the potential for alternative schemes that would provide the same end result equally well or better). However, there are clear ways in which the present scheme can and should be improved upon to ensure continuity between subsequent years. \textbf{We recommend that documentation be produced to guide future advocacy effort leaders in the planning of annual activities.} In practice, this documentation exists in a fragmented state, but the community would benefit from well-documented, commonly accessible resources. Additionally, \textbf{we recommend that documentation be produced to describe the technical aspects of the tools used to plan advocacy efforts (\textit{i.e.} WHIPS)}, in particular noting which aspects of maintenance need to be revisited each year to keep everything up-to-date and running smoothly (\textit{e.g.} periodically re-evaluating the numerical weight given to various members of Congress based on their committee assignments).

\subsection{Expanded Time Frame for Advocacy Activities} 

As discussed throughout this document, existing community advocacy efforts center on an annual, in-person, trip to Washington, D.C. in the Spring, generally timed to coincide with the drafting of appropriations and authorizations at the subcommittee level in Congress. There is some precedent for additional, targeted, in-person advocacy in Washington, D.C. by a smaller delegation -- in these cases the trip is timed to coincide with debate surrounding final details of the budget, generally in the Fall. It has repeatedly been observed by community members that there is opportunity to expand our model for advocacy to regularly include such ``off-cycle'' activities. This could also include, for example, trips to district offices (rather than Washington, D.C. offices) in the Fall to reinforce the messaging that is delivered each Spring.

Expanding the time frame in which the community undertakes advocacy would provide the opportunity to engage a broader fraction of the community in advocacy efforts, albeit at the expense of additional resources required on the organizational/management side. We do note that such an effort would clearly have been prohibitively resource-intensive before the advent of WHIPS, and that WHIPS potentially lowers the threshold for introducing additional advocacy activities to an acceptable level. One idea which has been suggested is to create a new profile tier within WHIPS to provide community members with the resources (\textit{e.g.} auto-generated letters) to easily engage in advocacy of only their local offices.

It is worth careful consideration of the marginal increase in resources required to sustain additional advocacy efforts throughout the year beyond what is already invested to support the current model of advocacy. We do not offer a recommendation as to whether such a model of more continuous advocacy is appropriate, though \textbf{we do recommend that a conversation around this topic be facilitated with more careful consideration of the potential costs and benefits.}

\subsection{HEP Advocacy technical tools/logistics}

\subsubsection{Washington High Energy Physics Integrated Planning System}

As noted above, the development of the Washington High Energy Physics Integrated Planning System (WHIPS) has been enormously additive to the annual advocacy efforts undertaken by the community. \textbf{We strongly endorse the continued utilization of this valuable resource, and recommend that steps be taken to ensure that its maintenance and administration be supported in future years.} Specifically, we note that the past and present developers of WHIPS are all early-career members of the community without permanent (long-term) positions -- this presents the potential for expertise to become unavailable if developers move out of the field as their careers progress. It has been suggested that community members with permanent positions that can become involved in the maintenance and administration be intentionally brought into the development team. Along the same line of reasoning, it has been proposed that the administration and development of WHIPS be formally brought under the ownership of one of the organizations that has historically led community advocacy efforts (or a future organization created more specifically for the purpose of leading advocacy efforts). Presently the monetary costs for upkeep of WHIPS, exclusive of labor, are provided by the Fermilab UEC and support the hosting of the website and database. While these costs are relatively minor, it is absolutely critical to the success of the ``DC Trip'' that these resources be continuously maintained.

We note that the developers of WHIPS maintain a list of technical features and functionalities that have been proposed by previous users, and observe that action on these recommendations has historically been focused on the imminent needs of advocacy efforts in a particular year. \textbf{We recommend that labor resources be allocated to continue development of WHIPS between advocacy cycles to better accommodate less-urgent, but still worthwhile, development tasks.}

\subsubsection{Grants Database, Other}

The annual advocacy effort has historically utilized a number of other technologies to organize and execute activities each year. Two of particular note are the ``wiki'' page, maintained by the Fermilab UEC, and the HEP funding webpage/grants database \cite{baumer}, originally developed by Michael Baumer. \textbf{We recommend that a long-term plan be developed for the continued utilization of these resources and that this plan include contingencies should the underlying technologies supporting these resources become unavailable.} We also note recent progress made to import aspects of the grants database into WHIPS.

Specific ideas which have been raised for how to expand the utility of the grants database include the following.

\begin{itemize}
    \item \textbf{Track funding by district.} This would enable participants of the annual advocacy effort to study funding trends for districts to which they are assigned. We note that this is a feature already in development within WHIPS through the integration of the grants database.
    \item \textbf{Develop a summary of NSF funding or details.} The funding breakdown is presently provided in much greater detail for grants provided by DOE OHEP than for grants provided by the NSF Division of Mathematical and Physical Sciences. We note that this asymmetry arises from the format of readily available data provided by the funding agencies, and that it may be difficult or impossible to analyze these data with finer resolution.
    \item \textbf{Develop an explicit vision for the future of this resource.} Previous participants of the advocacy effort have found this resource to be valuable and suggested specific ideas for how to improve it (including the above suggestions). \textbf{We recommend that a long-term plan for this resource be discussed between the developers and community advocacy leaders.} Actions that need to be taken sooner, rather than later, to support this future vision should be identified promptly.
\end{itemize}

\subsection{HEP Advocacy Diagnostics}

\subsubsection{Diagnostics to Characterize Impact of HEP Advocacy}

A repeated theme of suggestions made by community members that have participated in advocacy efforts in recent years is the development of diagnostics to understand the efficacy in engaging in advocacy as a community. The advent of WHIPS has made this much more technically feasible, and indeed certain diagnostics are presently built into the WHIPS database system (\textit{e.g.} tracking the advocacy activities of individual community members and for individual Congressional offices). However, we note that such questions are, by definition, outside the scope of the planning for any individual years' advocacy activities, due to the long time-scale on which the development of diagnostics would yield useful data. Therefore, \textbf{we recommend that a dedicated conversation be undertaken on this topic and that an intentional effort be made to expand the tracking of diagnostic information, within WHIPS or elsewhere.}

Examples of diagnostic information pertaining to the offices that we visit which could be tracked include:

\begin{itemize}
        \item Voting records of individual offices or Members of Congress, and
        \item Signatories of the ``Dear Colleague'' letters. 
\end{itemize}

\subsubsection{Diagnostics to Characterize Participation in HEP Advocacy}

We note that there may be value in tracking diagnostic information pertaining to the subset of our community that engages in advocacy. A recurring question during CEF06 conversations has been if the subset of the community that participates in advocacy activities is representative of the community at large. There is not comprehensive demographic information available for participation in the ``DC Trip'', especially farther back than recent years, but there is \textit{some} information available from post-trip surveys in recent years.

The groups organizing the annual advocacy effort (UEC, SLUO, USLUA, APS DPF) provide broad, but not complete, academic representation for the U.S. HEP community. For example, theorists make up a smaller fraction of the users communities than they do of the whole HEP community because the national laboratories focus largely on experimental efforts. From past survey results we also know that the majority of participants in the ``DC Trip'' have represented the Energy and Neutrino Frontiers, with smaller representation from Theory, Cosmic, Computational, and Instrumentation Frontiers (as defined in the Snowmass 2021-2022 process). According to the last few years of data, approximately one third of participants are postdocs, one third of participants are professors or national laboratory scientists, and the last third is distributed between Ph.D. students and other university or laboratory staff. It is expected (and indeed, intended) that the demographics of the group will skew towards early career members of the community, but more could be done to increase the representation from graduate students in particular. \textbf{We recommend that the tracking of these types of demographic categories continue and that efforts be undertaken as soon as possible to understand how inclusive present advocacy activities are across these categories.}

Historically, participation by underrepresented groups in HEP community advocacy activities has not been explicitly tracked. More effort could be done to quantify and correct representation through invitation of individuals or by working with groups that represent underrepresented demographics in HEP and in physics more broadly. We note that certain demographic categories are sensitive in nature and may not be relevant, and therefore \textbf{recommend that such discussions include the advice of DEI experts within our community.}

We note that one possibility to address the question of representation in community-driven advocacy efforts would be the creation of a new organization, separate from the three users groups, aimed only at HEP advocacy efforts. Through the explicit inclusion of other groups that are currently under-represented (\textit{e.g.} theorists), this body could achieve greater reflection of present and future demographics of the HEP community.

\subsubsection{Soliciting Feedback from Congressional Staff}

One suggestion that arose during the proceedings of CEF06 is the possibility of reaching out to contacts within Congressional offices and other organizations (\textit{e.g.} APS, AIP, AAAS) to solicit feedback and suggestions for how community advocacy efforts can be improved. There is anecdotal evidence from interactions between HEP community members and Congressional staff that most staffers won't even look at a long packet that is given to them in a meeting, and that if the goal is to have them read anything the better approach would be to instead only provide them with a single-page document. Perhaps, for example, the full packet described in Section \ref{sec:comm_materials} can be used to guide conversations in meetings, but only such a single-page summary document would be left with the office. It has also been suggested that such ``friendly'' contacts with policy expertise could serve as a resource for workshopping materials to understand if \textit{e.g.} they are pitched at the correct level of technical detail.

\subsubsection{Scientific Workforce Analysis}

One of the recommendations from the Snowmass 2013 report \cite{snowmass13recs} was to develop a sustainable process for collecting statistics on workforce development and technology transfer. The motivation behind this recommendation is that performing this scientific workforce analysis would help the community understand how much funding for early career members of the community comes from where. Such an investigation could also provide quantitative metrics reporting where early career community members funded by DOE go upon leaving the field. The availability of such data would strengthen one of the key messages utilized in community advocacy activities, which is of the impact of HEP on the overall economy as a driver for U.S. technological developments and innovation. DOE OHEP has started this effort and is interested in pursuing it but might need additional resources.

\subsection{Community Awareness about Advocacy Efforts}

We observe that awareness within the community of past and present advocacy efforts is highly heterogeneous, and assert that \textbf{work should be undertaken to increase awareness as well as to evaluate the present state of awareness.}

\subsubsection{Summarizing the DC trip each year}

\textbf{We recommend that a summary of the HEP advocacy activities of any particular year be assembled in a document suitable for distribution within the community.} We note that a high-level executive summary is presented each year to HEPAP, but the existence of this is not effectively broadcast within the community and it may not be accessible (\textit{i.e. understandable}) to members of the community that lack the public policy expertise of HEPAP at-large. We additionally note that certain details of the trip may not be suitable for consumption outside of the community (\textit{e.g.} themes of interactions with the offices of current Members of Congress), but believe that these details can be omitted with due care taken in preparation of a summary. The goal of such a summary should be to increase awareness of advocacy efforts in the community as well as interest in participation in future advocacy efforts. It has been noted that such a summary document could also include achievements from that year in the area of technology transfer. 

\subsubsection{Advocacy Training Materials and Support}

As part of annual advocacy efforts, training is provided to HEP community members that travel to Washington, D.C. on how to engage in advocacy for science. \textbf{We recommend that this training become more widely available within the community to increase the community's policy literacy overall, independently of the goals of any year's particular advocacy efforts.} This will additionally have the effect to raise awareness of community advocacy efforts within the HEP community.

We also note that in the course of CEF06 proceedings suggestions have been made for areas in which the training currently provided to community members can be improved. The prevailing theme among these suggestions is providing community members with more background on specific policy questions that are likely to be raised, including the details of policies currently in effect and policies currently being debated in Congress.

\subsubsection{HEP Community Communication Materials}

As discussed in detail in Section \ref{sec:dc_support}, the professional quality of the HEP community communication materials that are used for the annual advocacy effort is very impactful. These materials can be found at the US particle physics website \cite{uspp2022mar} and are jointly maintained by UEC, SLUO, and USLUA, and APS DPF. Before a specific effort was undertaken to improve the quality of these documents, the materials used during the annual advocacy effort came from many sources and were packaged into a plain blue folder. The messaging used in the new materials has not changed substantively, but the new materials are uniform in style, easy to read, and are aesthetically appealing, all of which have made the materials more impactful. \textbf{We strongly encourage continued support for producing these materials with a high level of quality.}

We note additionally that making these communication materials and guidance on how to effectively utilize them more broadly available within the HEP community would be beneficial. This would be specifically helpful in engagements between community members and legislators outside of the scope of community-organized advocacy, for example when Congressional staff visit experimental facilities (such as the national laboratories) and interact directly with researchers. Finally, we note that the quality and impact of these materials would benefit from a more systematic mechanism for feedback from community members. Increasing awareness of the existence of the materials would serve as a first step towards accomplishing this latter goal.
\section{Bigger-Picture Advocacy Questions}
\label{sec:big_picture_questions}

The following section will not focus on the HEP community's current efforts or on specifically actionable recommendations for how to change those efforts, but rather will take a broader look at science funding in the U.S. and potential changes that would benefit HEP and the wider scientific community.

HEP is in an opportune position to take a leading role in working with the U.S. science community to advocate for structural reform of government science funding. HEP advocacy efforts are well-organized and generally successful at achieving the basic aim of garnering support for basic research and for HEP. As practitioners of ``big science'', our community members have frequent interactions with funding agencies, which brings both insight into their concerns and experience managing large projects and budgets.

\subsection{Increasing R\&D Funding}

In order to address big-picture questions about how science is funded in the U.S., the scope of HEP advocacy would have to broaden considerably. Currently, the primary focus is on maintaining existing levels of funding with small increases to cover inflation and growth. While the overall funding level for HEP has increased since the previous Snowmass process in 2013~\cite{cef06paper3}, and this increase has helped the program in some ways, it can primarily be understood as a limited adjustment within the existing U.S. scientific funding infrastructure.

\begin{figure}[htb!]
\centering
\twofigeqh{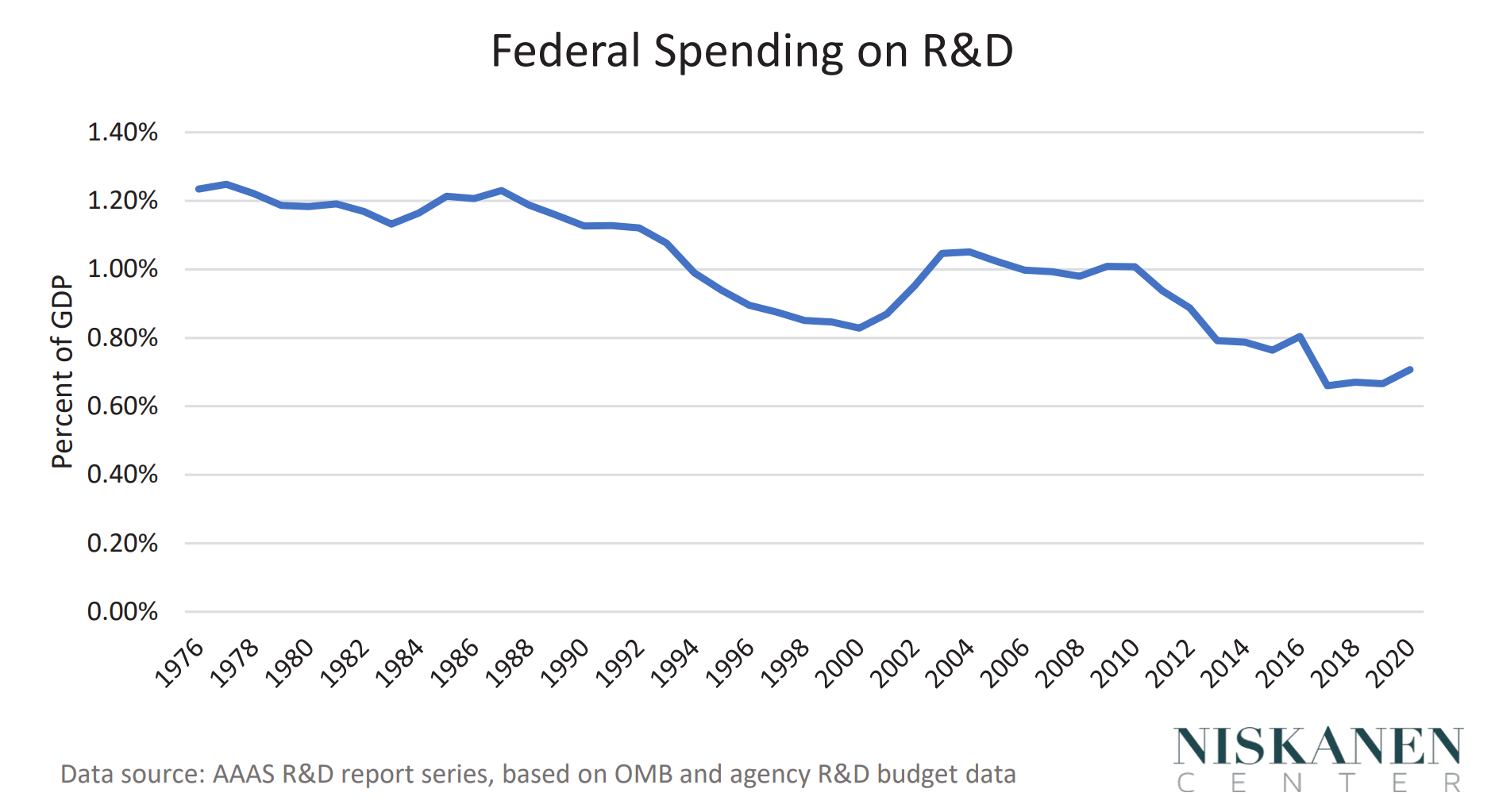}{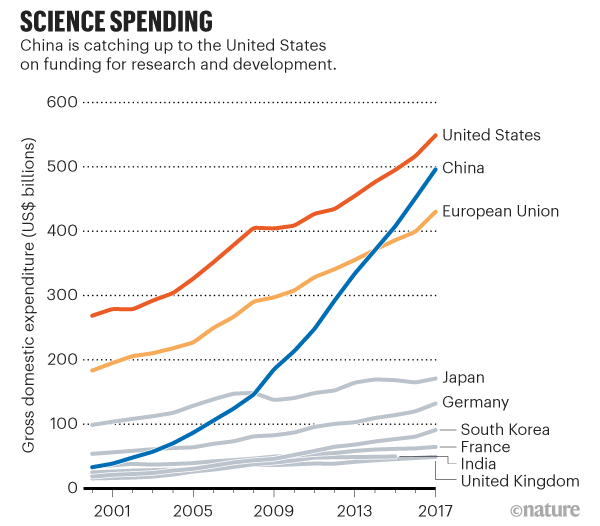}
\caption{Left: federal spending on R\&D as a percentage of GDP over time from Ref.~\cite{niskanen}.
Right: Comparison of science spending over time for the U.S., E.U., China, and other countries, from Ref.~\cite{nature}.
}
\label{fig:gdp_china}
\end{figure}

The case for reforming and expanding federal funding of science in the U.S. rests, first and foremost, on the two trends shown in Fig.~\ref{fig:gdp_china}. First, federal spending on R\&D has been decreasing as a percentage of GDP for four decades. Second, China has vastly increased its growth in spending in recent decades and now challenges the U.S. as the top spender, despite having a GDP only ${\sim}75\%$ that of the U.S. Therefore, it can be concluded that science spending in the U.S., despite growing in absolute terms, has not kept up with economic growth or international trends.

While it is broadly discussed that R\&D spending brings positive long-term economic returns, the full import of this fact is not reflected in current government science funding levels. Numerous studies have calculated positive rates of return ranging from 20\% to nearly 100\%, depending on various economic assumptions~\cite{SALTER2001509,summers}. By most estimates, the returns from scientific research are substantially higher than from private investments such as the stock market. In particular, Ref.~\cite{summers} additionally calculates the marginal rate of return from additional science spending above the current baseline and finds a positive value in all scenarios considered, with the majority greater than 50\%. These studies indicate that the U.S. could realize substantial economic and social gains by increasing science funding well above present levels.

The importance of truly basic research into new ideas with no known practical application cannot be overstated. It cannot be known what new science remains to be discovered nor what ideas will prove essential to solve future challenges, as the history of HEP illustrates. For example, the pioneers of quantum theory at the turn of the $20^{\mathrm{th}}$ century could not have predicted their work would lead to a deeper understanding of semiconductors, spawning an industry that has now grown to half a trillion dollars in yearly revenue from the nanometer-scale transistors found in every computer chip. The development of the World Wide Web at CERN to facilitate exchanges of data among physicists, in a similarly unpredictable fashion, led to massive economic growth from e-commerce and other large web-based industries.

Private industry is capable of investing sufficient capital into the ``development'' side of R\&D, as evidenced by the dramatic increase in such spending over the past few decades~\cite{increase2021}. Many technological developments result, directly or indirectly, from HEP research, as discussed in Section~\ref{sec:dc_support}. The federal government, because it aggregates large amounts of funding specifically not for profit-motivated activities, will always be a necessary supporter of basic research. Unfortunately, federal R\&D spending on basic research has actually decreased in the past decade~\cite{Mills2015}.

\subsection{The Granting Process}

There are other important structural improvements that can be made to U.S. science funding, beyond increasing the total amount of funding. In particular, the current approaches used by federal agencies to allocate funding via competitive grant proposals have been shown to be inefficient.

\subsubsection{Grant and Research Issues}

\begin{figure}[htb!]
\centering
\includegraphics[width=0.49\linewidth]{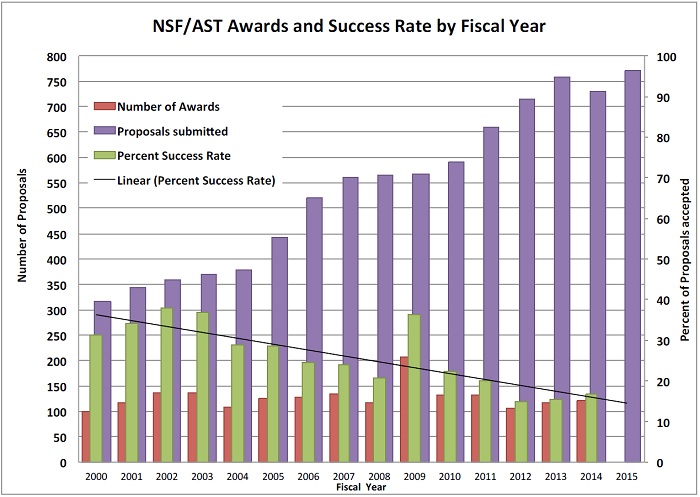}
\caption{The number of awards, number of proposals, and success rate since 2000 for NSF, from Ref.~\cite{Cushman2015}.
}
\label{fig:success_rate}
\end{figure}

A large survey of professors at research universities found that grant writing by itself consumes more than 4 hours per week on average, almost 10\% of a standard work week~\cite{LINK2008363}. Including other grant-related tasks such as administration increases this to nearly 20\%, and excluding professors without active grants to focus on principal investigators increases this further to nearly 40\%~\cite{Bozeman2015}. This level of administrative burden has been repeatedly confirmed by the National Science Board~\cite{workload2014reducing}, as well as the Federal Demonstration Partnership, which finds a similar value of 44\%~\cite{fdp2018}.
73\% of the professors surveyed in Ref.~\cite{Bozeman2015} highlighted the pressure and administrative burden of preparing grant proposals as the one thing they would change about their jobs. The total cost, in terms of researchers' and reviewers' time and effort, from preparing, submitting, and reviewing proposals can amount to a substantial fraction of the total budget for a given funding opportunity, in some cases even exceeding it~\cite{Gross2019}. This conclusion is observed in other countries with similar grant-based research programs, not just the U.S.~\cite{Herbert2013,nserc}. This inefficiency is compounded by the low and decreasing success rate for proposals: at NSF, this has decreased from above 30\% to below 20\% in recent decades, as depicted in Fig.~\ref{fig:success_rate}, with similar decreases for DOE, NIH, and other funding agencies~\cite{Cushman2015}.\\

Further, grant proposal reviews in particular have been found to be inaccurate and biased by numerous studies whose results are summarized here. There are poor correlations between reviewer scores and research success or productivity~\cite{Cole1981,Graves2011,Fang2016,Pier2018}. Women and members of underrepresented or minority groups are less likely to have their proposals funded, due to conscious or unconscious bias among reviewers~\cite{Ginther2011,Tabak2011,Witteman2019}. Reviewers are also biased against early-career researchers, whose success rates are correspondingly reduced to 10--15\% or even lower~\cite{impact2007,Daniels2015,DOE2019}. Instead, reviewers tend to give more funds to researchers who are already funded, which decreases the marginal impact of new research dollars~\cite{Lauer2017}. In addition, competitive peer-reviewed grant proposal systems may incentivize dishonest or unethical behavior~\cite{Conix2021}.

Beyond the reverse ageism and incumbency biases noted above, it has also been shown that reviewers systematically penalize more novel proposals~\cite{Boudreau2016,Nicholson2012}. This contributes to increasing conservatism and incrementalism throughout scientific fields. This trend has myriad negative effects: it decreases the long-term returns from basic research; it demoralizes or discourages scientific researchers with new ideas; and it exacerbates the overall stagnation and decline in innovation observed in recent decades~\cite{Bhattacharya2020}.

\subsubsection{Grant and Research Reform}

Given the numerous limitations and deficiencies in peer-reviewed competitive grant proposals described here, alternative systems should be considered. There has been popular discussion of extending the ARPA approach to other scientific fields. ARPA gives program managers broad latitude to distribute funding to potentially novel or risky ideas. While ARPA has had success in specific, limited contexts, it has disadvantages -- namely, its process is centralized and lacks guarantees of increasing novelty and extending the scope of scientific research. In addition, reform of the existing funding agencies should be preferred to appending a new system onto the existing systems.

The simplest possible system is described in Ref.~\cite{Vaesen2017}: equally distributing the entire pool of funding among all eligible researchers. This system would eliminate the majority of administrative costs from reviewing grants, which would increase the amount of funding available for research, in the optimistic assumption that the same total would be appropriated. It is estimated that each researcher, defined as any university faculty who engages in research, would receive almost \$600,000 to cover a five-year period. In comparison, the DOE Early Career Award provides \$750,000 over a five-year period for university researchers, after a months-long proposal preparation process and a similarly long review process, with an acceptance rate of only 10\%. (A more inclusive calculation of the acceptance rate, counting those who are discouraged from applying by the burdensome process and low published acceptance rate, would return an even smaller acceptance rate.) Clearly, many scientists, especially early career researchers, would benefit from such a system. It could be argued that such a flat system would encourage more egalitarian collaboration, since scientists would need to pool resources to accomplish larger tasks, and all scientists would arrive with equal contributions. However, it is also likely in practice that it would become difficult to pursue more costly research, such as the large projects found in HEP.

To allow larger grants while still avoiding issues that arise from a competitive, review-based approach, lotteries are being seriously considered by many nations and other organizations~\cite{Avin2015,Fang2016lottery,Adam2019}. While the introduction of explicit randomness may seem surprising, objective analyses of review-based grants, described above, show that the existing process is already largely random, and the non-random components reflect harmful biases at least as often as actual merit. The advantage of a completely random decision is that researchers can greatly reduce the amount of time they spend in preparing proposals, since the contents of the proposal are known to have no influence in the decision. Agencies would similarly reduce the time they spend in reviewing proposals. In practice, proposals would still be screened to meet a minimum standard, and the acceptance rate could be tuned to ensure that, on average, most researchers would have a successful proposal every few years. Ref.~\cite{Gross2019} describes in some detail how a competition-based approach could be modified to recover some of the efficiency gains from lotteries.

Another adjustment to the simplest flat system is a fully distributed approach, in which the allocation of funds is left to the entire corps of scientific researchers. In this method, each scientist would receive their flat allocation of funding, but would then be required to ``donate'' some minimum fraction of it to other researchers. Other requirements on donations may be imposed to prevent abuse and minimize bias. One version of this system called ``self-organized funding allocation'' (SOFA)~\cite{Bollen2014,Bollen2016,Bollen2019} has been proposed as a pilot program in the Netherlands.

A frequently discussed approach, often called ``fund people, not projects'', addresses the observation, discussed in more detail in Section~\ref{subsec:projectification}, that scientific research careers are indefinite rather than oriented toward fixed-term goals. SOFA has this property to some extent, but this approach can be implemented in other ways. Some federal funding agencies have specific programs that purport to behave this way, such as the DOE Early Career Award or the NIH Director's Pioneer Award. However, these programs have very low acceptance rates, high overall levels of bureaucracy, and their reviewers still often focus on the details of the specific research being proposed. The ``fund people'' approach is favored by a number of non-governmental research funding organizations, which often position themselves deliberately as alternatives to the existing system. The Howard Hughes Medical Institute~\cite{hhmi} is one of the oldest and most well-known entities to use such a system. An exhaustive review of its outcomes in comparison to NIH programs in Ref.~\cite{Ricon2020} indicates that it is generally successful, but it is unclear whether it could scale up to the entirety of U.S. R\&D.

A new idea called the bootstrap approach was discussed during the proceedings of CEF06. In this system, any scientist with a short proposal passing a minimum level of quality would be automatically eligible for a small amount of funding, such as \$20,000 for one year. Scientists who demonstrated success at the lower level of funding would automatically gain access to a higher tier of funding. Alternatively, scientists wishing to advance directly to a higher level could request a more stringent quality review. Success would be defined to include null results, in order to prevent publication bias. This method would allow new researchers to build up their expertise, laboratories, and working groups, while limiting the perceived risk to funding agencies from ``failing'' projects. The idea is conceptually related to prize-based funding, in which success brings additional resources for further research, as well as recognition. Prize-based approaches are often considered to be underutilized by economists and have been found to increase productivity~\cite{Jin2021}.

The recent COVID-19 pandemic has brought increased public attention to the persistent issues in science funding. In particular, it was revealed that Katalin Karik\'o, one of the key researchers in the development of the mRNA vaccines that objectively saved more than a million human lives, had her proposals repeatedly rejected and was denied tenure by her university~\cite{Garde2020}. This has led to the creation of several new non-governmental funding organizations, including Fast Grants~\cite{fastgrants}, the Arc Institute~\cite{instarc}, Arcadia Science~\cite{arcadia}, and New Science~\cite{newscience}. While private research institutions have always existed, this new group is explicitly aimed at addressing the deficiencies of current governmental funding approaches. The organizations listed above have all been funded at the level of tens to hundreds of millions of dollars, primarily raised from investors and private individuals. This impressive level of funding is likely to result in real impacts in both scientific results and science funding practices. However, it remains to be seen whether this level of private funding can be sustained in the long term.

U.S. federal funding agencies should investigate alternative systems with the goal of deploying pilot programs to understand their efficiency and viability. The alternative systems should not be limited to the examples discussed here. Pilot programs would enable a gradual rather than sudden approach to reform, which would hopefully broaden social and political acceptance of such changes. Despite the promising growth of non-governmental research funding, government funding remains crucial for the future of U.S. scientific leadership.

\subsection{Impact of Increasing ``Projectification'' and Targeted Funding}\label{subsec:projectification}

\begin{figure}[htb!]
\centering
\includegraphics[width=0.49\linewidth]{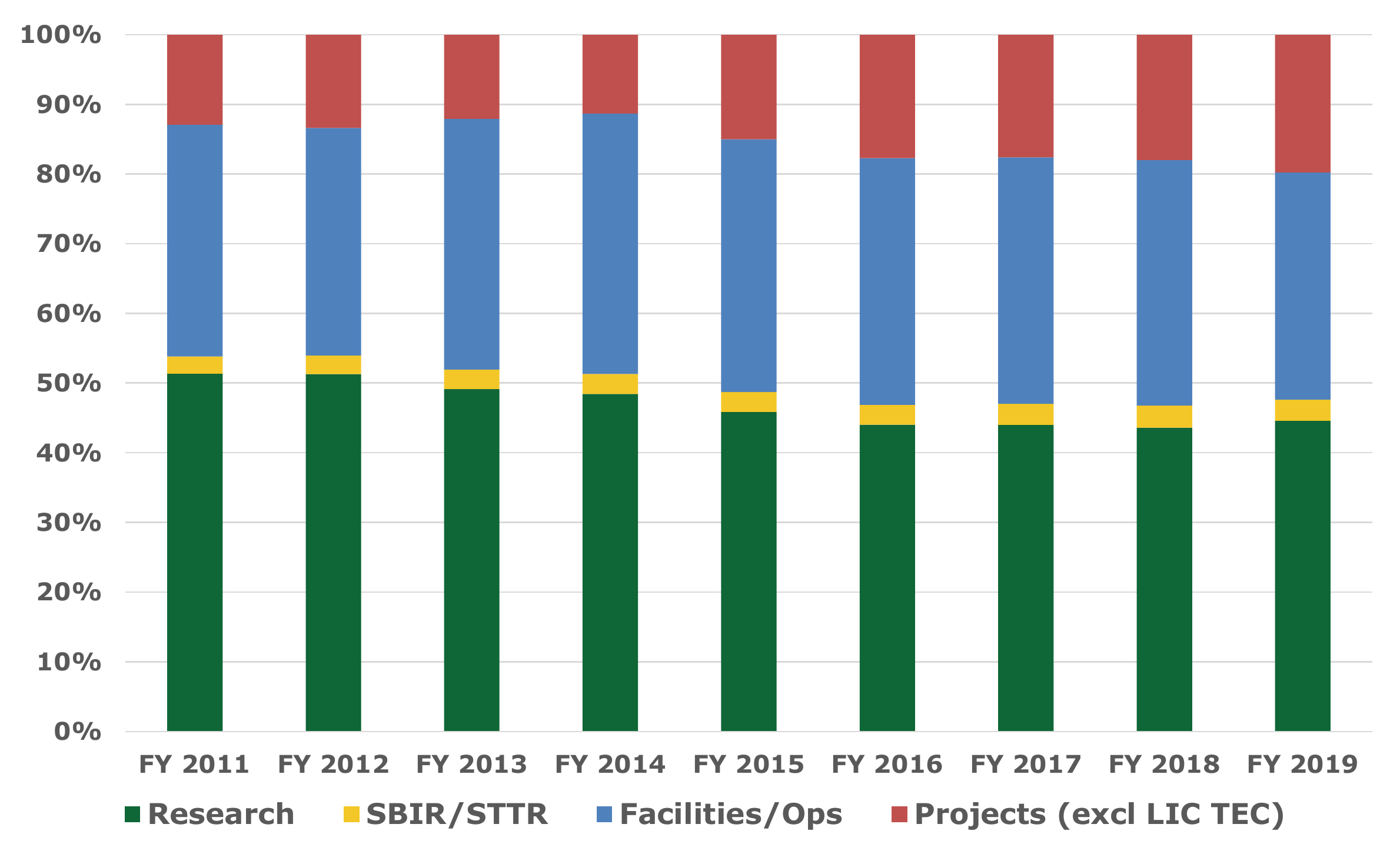}
\caption{The proportion of DOE Office of Science funding per year allocated to research, SBIR/STTR (Small Business Innovation Research/Small Business Technology Transfer), facilities/operations, and projects, from Ref.~\cite{nov_hepap}.
}
\label{fig:research}
\end{figure}

Building and operating facilities such as colliders and detectors whose costs range from many millions to billions of dollars requires careful project management and risk assessment. The large scale of these modern particle physics facilities has been shown to lead to corresponding increases in bureaucracy, conservatism, and incrementalism. The success of the LHC demonstrates the benefits of such an approach. However, the spread of this approach to other parts of HEP research, which may be called ``projectification'', can be shown to discourage novelty. For example, rejecting a young researcher's proposal because their ideas are too new or they have not had the resources to develop sufficient contingencies is harmful to the goal of scientific advancement. A project and risk management approach may not be necessary for all proposals, such as investigating a new theory of physics beyond the standard model or a new technique for data analysis. Further, the focus on large-scale projects has come at the expense of non-project funding for basic research, as shown in Fig.~\ref{fig:research}. In fact, from 2016 to 2019, core research funding in HEP faced a \$21M shortfall in real terms (considering inflation). Accounting for the \$27.5M of new spending on quantum information from 2017--2019, which is classified as core research, exacerbates the shortfall~\cite{nov_hepap}. This forces HEP scientists to modify their programs and proposals to fit available funding.

The past few years have seen numerous proposals in Congress to increase science funding, variously called the Endless Frontier Act, the United States Innovation and Competition Act, the DOE/NSF for the Future Acts, and the America Creating Opportunities for Manufacturing, Pre-Eminence in Technology, and Economic Strength Act. While each bill differs in its details, they all broadly focus on injecting new funding into very specific research areas that are named as national priorities. Some bills, more than others, additionally propose extending funding for basic research outside of those limited areas. However, the history of targeted research funding shows that it often fails to achieve its goals because of competing incentives between funding agencies and researchers. Ref.~\cite{Hall2021} describes how funding from the National Nanotechnology Initiative launched in 2000 did not lead to revolutions in nanotechnology but rather was captured by existing researchers to continue to fund their ongoing programs.

In the current system, if the focus of Congress and other government agencies is primarily on specific areas of interest, then funding for other areas can be expected to gradually decline in real terms in lieu of pressure to deliberately increase it. Projectification has been stated to lead funding agencies to treat research like a ``finite game''~\cite{Carse2012}: a bounded activity with a pre-established goal. While individual experiments and projects are indeed finite games, a scientific career is an ``infinite game'': the goal is to continue doing research indefinitely. Establishing a group and/or laboratory that can effectively conduct research and other scientific investigations is effort-intensive and time-consuming and involves building a repository of tacit, experiential knowledge that is not easily transferred. Governmental funding policies and procedures should be developed that take this into account explicitly, in order to avoid inefficiencies that arise from either scientists pivoting every few years with changes in funding priorities or scientists rephrasing their existing programs to capture new funding. One potentially helpful policy would be to require that any new targeted funding must be matched dollar for dollar by new non-targeted funding, as a way to achieve both national priorities and continue to support basic research. More generally, such sweeping bills should reflect substantial input from and consultation with the scientific community.

\subsection{Authorizations vs. Appropriations}

\begin{figure}[htb!]
\centering
\includegraphics[width=0.65\linewidth]{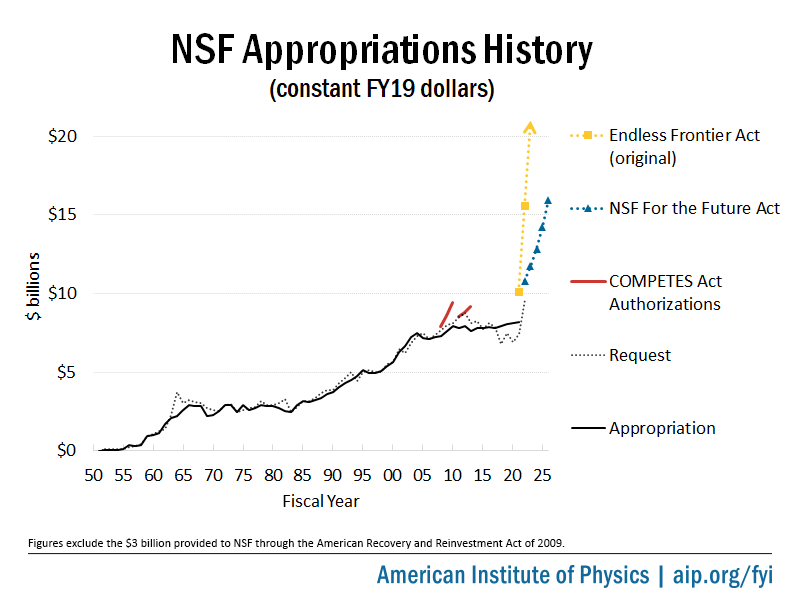}
\caption{A comparison of authorizations, requests, and appropriations for NSF~\cite{Ambrose2021}.
}
\label{fig:auth_approp}
\end{figure}

The scope of the recent bills for science funding listed in Section~\ref{subsec:projectification} must also be properly understood. These bills increased the amount of funding that Congress is \textit{authorized} to provide, but they do not require that Congress actually \textit{appropriates} that level of funding. Indeed, appropriations often do not match authorized levels, as shown in Fig.~\ref{fig:auth_approp}. It is vital that national attention to science funding does not end when authorization bills are passed, or the necessary funds will not be appropriated and real growth and improvements will not occur. Reform to the rest of the funding process is necessary in order to take advantage of any new authorizations.

We also remark here that the annual appropriation process itself, regardless of authorizations, significantly hinders U.S. leadership in science, including but not limited to HEP. Large-scale scientific experiments and projects are often not located in the U.S.; instead, the U.S. contributes funding and personpower to international efforts located on other continents. Attempts to establish similar international collaborations based in the U.S. have faced difficulties. One important contribution to this trend is the omnipresent possibility for Congress or funding agencies to reduce funding for a large-scale effort at any time. In other systems, such as those found in Europe, multiple years of funding can be appropriated and guaranteed at one time~\cite{Dalen2014,Abbott2019}. This helps build confidence in a large, collaborative project and assures collaborators of the funders' reliability. The U.S. should consider serious revisions in the budget appropriation process to facilitate projects in order to maintain our historical leadership.

\subsection{Congressional View of DOE-Supported Science} 

The Congressional debate on recent science authorization bills highlights another issue specific to funding for the physical sciences. The DOE Office of Science's most recent budget (FY22) was \$7.5 billion, at the same level as NSF's entire budget of \$8.8 billion. Most relevantly for this document, DOE funds 85\% of HEP research in the U.S. However, the national discussion of science funding most often includes only NSF and NIH, while DOE's sizable contributions are ignored. DOE was entirely excluded from the initial draft of the Endless Frontier Act, and the co-sponsoring Senator initially described its later addition in an amendment as a ``poison pill''~\cite{Ambrose2021may} before eventually acknowledging DOE's substantial contributions to science, often through its national laboratories~\cite{Ambrose2021jun}. It is important for HEP that recognition of the DOE Office of Science increases until it is routinely mentioned along with NSF in science funding discussions.

\subsection{Summary of Discussions on Bigger-Picture Advocacy Questions}

The bigger-picture questions discussed throughout Section \ref{sec:big_picture_questions} touch not only on HEP but on all science in the U.S. The matters discussed have scopes reaching far beyond HEP, but the benefits from acting on these topics could be substantial to the HEP community. And, they relate to all areas of HEP community government engagement, including interactions with Congress as well as with the funding agencies and other executive branch agencies. The specifics of existing and proposed future HEP community advocacy of the funding agencies and other executive branch agencies are discussed in detail in Ref.~\cite{cef06paper3}.\\

\noindent On federal spending on scientific R\&D, we note:

\begin{itemize}
    \item U.S. federal spending on scientific R\&D has decreased as a percentage of GDP;
    \item China is now challenging the U.S. as the top R\&D spender;
    \item R\&D spending brings large economic returns in the long run; and
    \item the marginal rate of return for additional R\&D spending is likely still high.
\end{itemize}

Increased U.S. federal R\&D spending will have positive benefits, and discussions about reforms and changes to U.S. federal R\&D policy should receive substantive input from the entire scientific community.\\

\noindent On grant and research funding reform, we observe:

\begin{itemize}
    \item the process of competitive grant proposals to distribute funding has been shown to be an inefficient use of scientists' time and agencies' efforts;
    \item peer review of competitive grant proposals is a source of numerous biases, including against women, members of minority groups, early-career researchers, and novel ideas, and in favor of top scientists who are already adequately funded and for whom the marginal impact of additional funding is small;
    \item grant proposal success rates have decreased significantly in recent decades; and
    \item private industry has increased its spending on the development side of R\&D funding.
\end{itemize}

The funding agencies should work to end bureaucracy, conservatism, and bias in the process for distributing funding to scientists. Alternate methods of distributing funding, such as flat distributions, lotteries, self-organized systems, or bootstrap systems, should be explored by funding agencies with pilot projects.\\

\noindent On ``projectification'', we discuss how:

\begin{itemize}
    \item the ``projectification'' of research funding further decreases novelty, especially in HEP;
    \item recent proposals to increase U.S. R\&D spending primarily target specific research topics; and
    \item targeted funding for specific research topics rarely results in major progress on those topics because of competing incentives.
\end{itemize}

Approaches should be considered to control the growth of projectification and identify which areas need to follow a project-based approach. Funding increases often go to projects or targeted areas; if targeted funding were matched dollar for dollar with non-targeted funding for basic research, this trend would be reversed.\\

\noindent On Congressional appropriations practices, we remark:

\begin{itemize}
    \item Congress typically appropriates less funding than authorized;
    \item the yearly appropriation process decreases international confidence in U.S.-based large-scale scientific projects, such as colliders; and
    \item the DOE was initially excluded from some recent authorization proposals, which is concerning for HEP.
\end{itemize}%

Funding authorizations should be more closely coupled to appropriations, to ensure that the authorized funding actually becomes available. Reform of the appropriations process to guarantee multiple years of funding for approved projects could increase international confidence in U.S.-led research. DOE should publicly emphasize the Office of Science's important contributions in order to ensure its funding grows along with any increases to NSF's funding.\\

\noindent A serious effort by the U.S. research funding apparatus to institute these improvements and reforms would be transformative for all fields, including HEP, and would ensure continuing U.S. leadership in scientific innovation.

\section{Conclusion}
\label{sec:conclusion}

This paper describes the annual advocacy efforts undertaken by the HEP community that pertain directly to the funding of our field, \textit{i.e.} the annual ``DC Trip''.  We have provided a detailed presentation of the current activities and the community communication material used in these efforts. We have discussed several actionable items that could be implemented as direct improvements to the current efforts, mostly without substantial new effort or resources. We have also introduced a bigger-picture discussion of the nature of governmental funding for HEP and have suggested actions that could be taken within the community or within Congress that would be beneficial to the field.

\bibliographystyle{unsrtnat}
\bibliography{Bibliography/common,Bibliography/main}

\begin{thebibliography}{76}
\providecommand{\natexlab}[1]{#1}
\providecommand{\url}[1]{\texttt{#1}}
\expandafter\ifx\csname urlstyle\endcsname\relax
  \providecommand{\doi}[1]{doi: #1}\else
  \providecommand{\doi}{doi: \begingroup \urlstyle{rm}\Url}\fi

\bibitem[Bardeen et~al.(2013)Bardeen, Cronin-Hennessy, White, and
  Yurkewicz]{snowmass13recs}
M.~Bardeen, D.~Cronin-Hennessy, H.~White, and K.~Yurkewicz.
\newblock {Communication with {U.S.} Policy Makers and Opinion Leaders}, 2013.
\newblock URL
  \url{https://www.slac.stanford.edu/econf/C1307292/docs/CommunicationEducationOutreach/PolicyMakers-51.pdf}.

\bibitem[Diurba et~al.(2022{\natexlab{a}})]{cef06paper2}
Richie Diurba et~al.
\newblock {Snowmass '21 Community Engagement Frontier 6: Public Policy and
  Government Engagement: Congressional Advocacy for Areas Beyond HEP Funding},
  2022{\natexlab{a}}.
\newblock URL \url{https://arxiv.org/abs/2207.00124}.

\bibitem[Diurba et~al.(2022{\natexlab{b}})]{cef06paper3}
Richie Diurba et~al.
\newblock {Snowmass '21 Community Engagement Frontier 6: Public Policy and
  Government Engagement: Non-Congressional Government Engagement},
  2022{\natexlab{b}}.
\newblock URL \url{https://arxiv.org/abs/2207.00125}.

\bibitem[Siegrist(2022)]{march_hepap}
Jim Siegrist.
\newblock {DOE presentation HEPAP March 2022}, 2022.
\newblock URL \url{https://science.osti.gov/hep/hepap/Meetings/202203}.

\bibitem[Baumer()]{baumer}
Michael Baumer.
\newblock {US HEP Grants Database}.
\newblock URL \url{https://mbaumer.github.io/us_hep_funding/}.

\bibitem[DC_()]{DC_trip_wiki}
{DC trip wiki (not publicly accessible)}.
\newblock URL \url{https://www.uec-whips.org/wiki/index.php/Main_Page}.

\bibitem[hep(2017)]{hepap_2017}
{2017 HEPAP meeting}, 2017.
\newblock URL \url{https://science.osti.gov/hep/hepap/Meetings/201706}.

\bibitem[hep(2018)]{hepap_2018}
{2018 HEPAP meeting}, 2018.
\newblock URL \url{https://science.osti.gov/hep/hepap/Meetings/201805}.

\bibitem[hep(2019)]{hepap_2019}
{2019 HEPAP meeting}, 2019.
\newblock URL \url{https://science.osti.gov/hep/hepap/Meetings/201905}.

\bibitem[hep(2020)]{hepap_2020}
{2020 HEPAP meeting}, 2020.
\newblock URL \url{https://science.osti.gov/hep/hepap/Meetings/202007}.

\bibitem[hep(2021)]{hepap_2021}
{2021 HEPAP meeting}, 2021.
\newblock URL \url{https://science.osti.gov/hep/hepap/Meetings/202111}.

\bibitem[usp(2022)]{uspp2022mar}
{U.S. Particle Physics: Building for Discovery}, March 2022.
\newblock URL \url{https://www.usparticlephysics.org/}.

\bibitem[{Pew Research Center}(2020)]{pew2020aug}
{Pew Research Center}.
\newblock {Election 2020: Voters Are Highly Engaged, but Nearly Half Expect To
  Have Difficulties Voting}, August 2020.
\newblock URL
  \url{https://www.pewresearch.org/politics/2020/08/13/important-issues-in-the-2020-election/}.

\bibitem[{Gallup}(2022)]{gallup2022mar}
{Gallup}.
\newblock Most important problem, March 2022.
\newblock URL
  \url{https://news.gallup.com/poll/1675/most-important-problem.aspx}.

\bibitem[{Congress.gov}(2020)]{naiia2020}
{Congress.gov}.
\newblock {H.R.6216 - 116th Congress (2019-2020): National Artificial
  Intelligence Initiative Act of 2020}, March 2020.
\newblock URL
  \url{https://www.congress.gov/bill/116th-congress/house-bill/6216}.

\bibitem[aig(2022)]{aigov2022}
{National Artificial Intelligence Initiative}, March 2022.
\newblock URL \url{https://www.ai.gov/}.

\bibitem[ost(2022)]{osti_qis_2022mar}
{Quantum Information Science (QIS)}, March 2022.
\newblock URL \url{https://science.osti.gov/Initiatives/QIS}.

\bibitem[{Congress.gov}(2019)]{ndaa2020}
{Congress.gov}.
\newblock {S.1790 - 116th Congress (2019-2020): National Defense Authorization
  Act for Fiscal Year 2020}, December 2019.
\newblock URL
  \url{https://www.congress.gov/bill/116th-congress/senate-bill/1790}.

\bibitem[nlj(2020)]{nljaa2020}
{S.1739 - 116th Congress (2019-2020): Department of Energy National Labs Jobs
  ACCESS Act}.
\newblock January 2020.
\newblock URL
  \url{https://www.congress.gov/bill/116th-congress/senate-bill/1739}.

\bibitem[fna()]{fnal_factsheets}
Fermilab fact sheets.
\newblock URL \url{https://news.fnal.gov/newsroom/fact-sheets-and-brochures/}.

\bibitem[sla()]{slac_factsheets}
Slac fact sheets.
\newblock URL \url{https://www6.slac.stanford.edu/about/fact-sheets}.

\bibitem[hig(2013)]{higgs}
{Higgs day on the Hill}, 2013.
\newblock URL
  \url{https://news.fnal.gov/2013/12/recognition-on-capitol-hill-of-u-s-participation-in-higgs-discovery/}.

\bibitem[aas(2022)]{aas_cvd_2022mar}
How {AAS} advocates, March 2022.
\newblock URL
  \url{https://aas.org/advocacy/how-aas-advocates/congressional-visits-days}.

\bibitem[aps(2022)]{aps_gov_affairs_2022mar}
Policy \& advocacy, March 2022.
\newblock URL \url{https://www.aps.org/policy/}.

\bibitem[Lindsey and Hammond(2020)]{niskanen}
Brink Lindsey and Samuel Hammond.
\newblock Faster growth, fairer growth: Policies for a high road, high
  performance economy, 2020.
\newblock URL
  \url{https://www.niskanencenter.org/faster-growth-fairer-growth-policies-for-a-high-road-high-performance-economy/}.
\newblock Niskanen Center.

\bibitem[Viglione(2020)]{nature}
Giuliana Viglione.
\newblock China is closing gap with {United States} on research spending.
\newblock 2020.
\newblock \doi{10.1038/d41586-020-00084-7}.
\newblock URL \url{https://www.nature.com/articles/d41586-020-00084-7}.

\bibitem[Salter and Martin(2001)]{SALTER2001509}
Ammon~J. Salter and Ben~R. Martin.
\newblock The economic benefits of publicly funded basic research: a critical
  review.
\newblock \emph{Research Policy}, 30\penalty0 (3):\penalty0 509--532, 2001.
\newblock ISSN 0048-7333.
\newblock \doi{https://doi.org/10.1016/S0048-7333(00)00091-3}.
\newblock URL
  \url{https://www.sciencedirect.com/science/article/pii/S0048733300000913}.

\bibitem[Jones and Summers(2020)]{summers}
Benjamin~F. Jones and Lawrence~H. Summers.
\newblock A calculation of the social returns to innovation.
\newblock \emph{NBER}, 2020.
\newblock \doi{10.3386/w27863}.

\bibitem[Boroush(2021)]{increase2021}
Mark Boroush.
\newblock {U.S.} {R\&D} increased by \$51 billion in 2018, to \$606 billion;
  estimate for 2019 indicates a further rise to \$656 billion.
\newblock 2021.
\newblock URL \url{https://ncses.nsf.gov/pubs/nsf21324}.

\bibitem[Mills(2015)]{Mills2015}
Mark~P. Mills.
\newblock Basic research and the innovation frontier: Decentralizing federal
  support and stimulating market solutions, Feb 2015.
\newblock URL
  \url{https://www.manhattan-institute.org/html/basic-research-and-innovation-frontier-decentralizing-federal-support-and-stimulating-market}.

\bibitem[Link et~al.(2008)Link, Swann, and Bozeman]{LINK2008363}
Albert~N. Link, Christopher~A. Swann, and Barry Bozeman.
\newblock A time allocation study of university faculty.
\newblock \emph{Economics of Education Review}, 27\penalty0 (4):\penalty0
  363--374, 2008.
\newblock ISSN 0272-7757.
\newblock \doi{https://doi.org/10.1016/j.econedurev.2007.04.002}.
\newblock URL
  \url{https://www.sciencedirect.com/science/article/pii/S0272775707000623}.

\bibitem[Bozeman(2015)]{Bozeman2015}
Barry Bozeman.
\newblock Bureaucratization in academic research policy: perspectives from red
  tape theory.
\newblock In \emph{20th International Conference on Science and Technology
  Indicators}, November 2015.
\newblock \doi{10.13140/RG.2.1.3496.5848}.

\bibitem[{National Science Board}(2014)]{workload2014reducing}
{National Science Board}.
\newblock Reducing investigators' administrative workload for federally funded
  research.
\newblock 2014.
\newblock URL \url{https://www.nsf.gov/pubs/2014/nsb1418/nsb1418.pdf}.

\bibitem[Schneider(2020)]{fdp2018}
Sandra~L. Schneider.
\newblock 2018 faculty workload survey, Sep 2020.
\newblock URL
  \url{https://thefdp.org/default/assets/File/Documents/FDP\%20FWS\%202018\%20Primary\%20Report.pdf}.

\bibitem[Gross and Bergstrom(2019)]{Gross2019}
Kevin Gross and Carl~T. Bergstrom.
\newblock Contest models highlight inherent inefficiencies of scientific
  funding competitions.
\newblock \emph{{PLOS} Biology}, 17\penalty0 (1):\penalty0 e3000065, January
  2019.
\newblock \doi{10.1371/journal.pbio.3000065}.
\newblock URL \url{https://doi.org/10.1371/journal.pbio.3000065}.

\bibitem[Herbert et~al.(2013)Herbert, Barnett, and Graves]{Herbert2013}
Danielle~L. Herbert, Adrian~G. Barnett, and Nicholas Graves.
\newblock Australia's grant system wastes time.
\newblock \emph{Nature}, 495\penalty0 (7441):\penalty0 314--314, Mar 2013.
\newblock ISSN 1476-4687.
\newblock \doi{10.1038/495314d}.
\newblock URL \url{https://doi.org/10.1038/495314d}.

\bibitem[Gordon and Poulin(2009)]{nserc}
Richard Gordon and Bryan Poulin.
\newblock Cost of the {NSERC} science grant peer review system exceeds the cost
  of giving every qualified researcher a baseline grant.
\newblock \emph{Accountability in research}, 16:\penalty0 13--40, 02 2009.
\newblock \doi{10.1080/08989620802689821}.

\bibitem[Cushman et~al.(2015)Cushman, Hoeksema, Kouveliotou, Lowenthal,
  Peterson, Stassun, and von Hippel]{Cushman2015}
Priscilla Cushman, Todd Hoeksema, Chryssa Kouveliotou, James Lowenthal, Bradley
  Peterson, Keivan Stassun, and Ted von Hippel.
\newblock Impact of declining proposal success rates on scientific
  productivity.
\newblock \emph{Astronomy and Astrophysics Advisory Committee}, 10 2015.

\bibitem[Cole et~al.(1981)Cole, Cole, and Simon]{Cole1981}
Stephen Cole, Jonathan~R. Cole, and Gary~A. Simon.
\newblock Chance and consensus in peer review.
\newblock \emph{Science}, 214\penalty0 (4523):\penalty0 881--886, November
  1981.
\newblock \doi{10.1126/science.7302566}.
\newblock URL \url{https://doi.org/10.1126/science.7302566}.

\bibitem[Graves et~al.(2011)Graves, Barnett, and Clarke]{Graves2011}
N.~Graves, A.~G. Barnett, and P.~Clarke.
\newblock Funding grant proposals for scientific research: retrospective
  analysis of scores by members of grant review panel.
\newblock \emph{{BMJ}}, 343\penalty0 (sep27 1):\penalty0 d4797--d4797,
  September 2011.
\newblock \doi{10.1136/bmj.d4797}.
\newblock URL \url{https://doi.org/10.1136/bmj.d4797}.

\bibitem[Fang et~al.(2016)Fang, Bowen, and Casadevall]{Fang2016}
Ferric~C Fang, Anthony Bowen, and Arturo Casadevall.
\newblock {NIH} peer review percentile scores are poorly predictive of grant
  productivity.
\newblock \emph{{eLife}}, 5, February 2016.
\newblock \doi{10.7554/elife.13323}.
\newblock URL \url{https://doi.org/10.7554/elife.13323}.

\bibitem[Pier et~al.(2018)Pier, Brauer, Filut, Kaatz, Raclaw, Nathan, Ford, and
  Carnes]{Pier2018}
Elizabeth~L. Pier, Markus Brauer, Amarette Filut, Anna Kaatz, Joshua Raclaw,
  Mitchell~J. Nathan, Cecilia~E. Ford, and Molly Carnes.
\newblock Low agreement among reviewers evaluating the same {NIH} grant
  applications.
\newblock \emph{Proceedings of the National Academy of Sciences}, 115\penalty0
  (12):\penalty0 2952--2957, March 2018.
\newblock \doi{10.1073/pnas.1714379115}.
\newblock URL \url{https://doi.org/10.1073/pnas.1714379115}.

\bibitem[Ginther et~al.(2011)Ginther, Schaffer, Schnell, Masimore, Liu, Haak,
  and Kington]{Ginther2011}
Donna~K. Ginther, Walter~T. Schaffer, Joshua Schnell, Beth Masimore, Faye Liu,
  Laurel~L. Haak, and Raynard Kington.
\newblock Race, ethnicity, and {NIH} research awards.
\newblock \emph{Science}, 333\penalty0 (6045):\penalty0 1015--1019, August
  2011.
\newblock \doi{10.1126/science.1196783}.
\newblock URL \url{https://doi.org/10.1126/science.1196783}.

\bibitem[Tabak and Collins(2011)]{Tabak2011}
Lawrence~A. Tabak and Francis~S. Collins.
\newblock Weaving a richer tapestry in biomedical science.
\newblock \emph{Science}, 333\penalty0 (6045):\penalty0 940--941, August 2011.
\newblock \doi{10.1126/science.1211704}.
\newblock URL \url{https://doi.org/10.1126/science.1211704}.

\bibitem[Witteman et~al.(2019)Witteman, Hendricks, Straus, and
  Tannenbaum]{Witteman2019}
Holly~O Witteman, Michael Hendricks, Sharon Straus, and Cara Tannenbaum.
\newblock Are gender gaps due to evaluations of the applicant or the science?
  {A} natural experiment at a national funding agency.
\newblock \emph{The Lancet}, 393\penalty0 (10171):\penalty0 531--540, February
  2019.
\newblock \doi{10.1016/s0140-6736(18)32611-4}.
\newblock URL \url{https://doi.org/10.1016/s0140-6736(18)32611-4}.

\bibitem[{National Science Foundation}(2007)]{impact2007}
{National Science Foundation}.
\newblock Impact of proposal and award management mechanisms.
\newblock 2007.
\newblock URL \url{http://www.nsf.gov/pubs/2007/nsf0745/nsf0745.pdf}.

\bibitem[Daniels(2015)]{Daniels2015}
Ronald~J. Daniels.
\newblock A generation at risk: Young investigators and the future of the
  biomedical workforce.
\newblock \emph{Proceedings of the National Academy of Sciences}, 112\penalty0
  (2):\penalty0 313--318, January 2015.
\newblock \doi{10.1073/pnas.1418761112}.
\newblock URL \url{https://doi.org/10.1073/pnas.1418761112}.

\bibitem[Cooke()]{DOE2019}
Michael Cooke.
\newblock High energy physics budget, early career, diversity and inclusion.
\newblock URL
  \url{https://indico.cern.ch/event/782953/contributions/3515782/attachments/1887846/3114603/2019-07-29_DOE_HEP_Budget_Early_Career_Diversity_and_Inclusion.pdf}.

\bibitem[Lauer et~al.(2017)Lauer, Roychowdhury, Patel, Walsh, and
  Pearson]{Lauer2017}
Michael Lauer, Deepshikha Roychowdhury, Katie Patel, Rachael Walsh, and Katrina
  Pearson.
\newblock Marginal returns and levels of research grant support among
  scientists supported by the {National Institutes of Health}.
\newblock May 2017.
\newblock \doi{10.1101/142554}.
\newblock URL \url{https://doi.org/10.1101/142554}.

\bibitem[Conix et~al.(2021)Conix, Block, and Vaesen]{Conix2021}
Stijn Conix, Andreas~De Block, and Krist Vaesen.
\newblock Grant writing and grant peer review as questionable research
  practices.
\newblock \emph{F1000Research}, 10:\penalty0 1126, December 2021.
\newblock \doi{10.12688/f1000research.73893.2}.
\newblock URL \url{https://doi.org/10.12688/f1000research.73893.2}.

\bibitem[Boudreau et~al.(2016)Boudreau, Guinan, Lakhani, and
  Riedl]{Boudreau2016}
Kevin~J. Boudreau, Eva~C. Guinan, Karim~R. Lakhani, and Christoph Riedl.
\newblock Looking across and looking beyond the knowledge frontier:
  Intellectual distance, novelty, and resource allocation in science.
\newblock \emph{Management Science}, 62\penalty0 (10):\penalty0 2765--2783,
  2016.
\newblock \doi{10.1287/mnsc.2015.2285}.

\bibitem[Nicholson and Ioannidis(2012)]{Nicholson2012}
Joshua~M. Nicholson and John P.~A. Ioannidis.
\newblock Conform and be funded.
\newblock \emph{Nature}, 492\penalty0 (7427):\penalty0 34--36, December 2012.
\newblock \doi{10.1038/492034a}.
\newblock URL \url{https://doi.org/10.1038/492034a}.

\bibitem[Bhattacharya and Packalen(2020)]{Bhattacharya2020}
Jay Bhattacharya and Mikko Packalen.
\newblock Stagnation and scientific incentives.
\newblock Technical report, February 2020.
\newblock URL \url{https://doi.org/10.3386/w26752}.

\bibitem[Vaesen and Katzav(2017)]{Vaesen2017}
Krist Vaesen and Joel Katzav.
\newblock How much would each researcher receive if competitive government
  research funding were distributed equally among researchers?
\newblock \emph{{PLOS} {ONE}}, 12\penalty0 (9):\penalty0 e0183967, September
  2017.
\newblock \doi{10.1371/journal.pone.0183967}.
\newblock URL \url{https://doi.org/10.1371/journal.pone.0183967}.

\bibitem[Avin(2015)]{Avin2015}
Shahar Avin.
\newblock Funding science by lottery.
\newblock In \emph{Recent Developments in the Philosophy of Science: {EPSA}13
  Helsinki}, pages 111--126. Springer International Publishing, 2015.
\newblock \doi{10.1007/978-3-319-23015-3_9}.
\newblock URL \url{https://doi.org/10.1007/978-3-319-23015-3_9}.

\bibitem[Fang and Casadevall(2016)]{Fang2016lottery}
Ferric~C. Fang and Arturo Casadevall.
\newblock Research funding: the case for a modified lottery.
\newblock \emph{{mBio}}, 7\penalty0 (2), May 2016.
\newblock \doi{10.1128/mbio.00422-16}.
\newblock URL \url{https://doi.org/10.1128/mbio.00422-16}.

\bibitem[Adam(2019)]{Adam2019}
David Adam.
\newblock Science funders gamble on grant lotteries.
\newblock \emph{Nature}, 575\penalty0 (7784):\penalty0 574--575, November 2019.
\newblock \doi{10.1038/d41586-019-03572-7}.
\newblock URL \url{https://doi.org/10.1038/d41586-019-03572-7}.

\bibitem[Bollen et~al.(2014)Bollen, Crandall, Junk, Ding, and
  B\"{o}rner]{Bollen2014}
Johan Bollen, David Crandall, Damion Junk, Ying Ding, and Katy B\"{o}rner.
\newblock From funding agencies to scientific agency.
\newblock \emph{{EMBO} reports}, 15\penalty0 (2):\penalty0 131--133, January
  2014.
\newblock \doi{10.1002/embr.201338068}.
\newblock URL \url{https://doi.org/10.1002/embr.201338068}.

\bibitem[Bollen et~al.(2016)Bollen, Crandall, Junk, Ding, and
  B\"{o}rner]{Bollen2016}
Johan Bollen, David Crandall, Damion Junk, Ying Ding, and Katy B\"{o}rner.
\newblock An efficient system to fund science: from proposal review to
  peer-to-peer distributions.
\newblock \emph{Scientometrics}, 110\penalty0 (1):\penalty0 521--528, September
  2016.
\newblock \doi{10.1007/s11192-016-2110-3}.
\newblock URL \url{https://doi.org/10.1007/s11192-016-2110-3}.

\bibitem[Bollen et~al.(2019)Bollen, Carpenter, Lubchenco, and
  Scheffer]{Bollen2019}
Johan Bollen, Stephen~R. Carpenter, Jane Lubchenco, and Marten Scheffer.
\newblock Rethinking resource allocation in science.
\newblock \emph{Ecology and Society}, 24\penalty0 (3), 2019.
\newblock \doi{10.5751/es-11005-240329}.
\newblock URL \url{https://doi.org/10.5751/es-11005-240329}.

\bibitem[hhm()]{hhmi}
{Howard Hughes Medical Institute}.
\newblock URL \url{https://www.hhmi.org/}.

\bibitem[Ric\'on(2020)]{Ricon2020}
Jos\'e~Luis Ric\'on.
\newblock Fund people, not projects {I}: The {HHMI} and the {NIH Director's
  Pioneer Award}, Dec 2020.
\newblock URL \url{https://nintil.com/hhmi-and-nih/}.

\bibitem[Jin et~al.(2021)Jin, Ma, and Uzzi]{Jin2021}
Ching Jin, Yifang Ma, and Brian Uzzi.
\newblock Scientific prizes and the extraordinary growth of scientific topics.
\newblock \emph{Nature Communications}, 12\penalty0 (1), October 2021.
\newblock \doi{10.1038/s41467-021-25712-2}.
\newblock URL \url{https://doi.org/10.1038/s41467-021-25712-2}.

\bibitem[Garde(2020)]{Garde2020}
Damien Garde.
\newblock The story of {mRNA}: How a once-dismissed idea became a leading
  technology in the {Covid} vaccine race, Nov 2020.
\newblock URL
  \url{https://www.statnews.com/2020/11/10/the-story-of-mrna-how-a-once-dismissed-idea-became-a-leading-technology-in-the-covid-vaccine-race/}.

\bibitem[fas()]{fastgrants}
{Fast Grants}.
\newblock URL \url{https://fastgrants.org/}.

\bibitem[ins()]{instarc}
{Arc Institute}.
\newblock URL \url{https://arcinstitute.org/}.

\bibitem[arc()]{arcadia}
{Arcadia Science}.
\newblock URL \url{https://www.arcadia.science/}.

\bibitem[new()]{newscience}
{New Science}.
\newblock URL \url{https://newscience.org/}.

\bibitem[Stone(2019)]{nov_hepap}
Alan Stone.
\newblock {High Energy Physics Budget Planning and Execution}, 2019.
\newblock URL \url{https://science.osti.gov/hep/hepap/Meetings/201911}.

\bibitem[Hall(2021)]{Hall2021}
J~Hall.
\newblock \emph{Where is my flying car?}
\newblock Stripe Press, Toronto, 2021.
\newblock ISBN 1953953182.

\bibitem[Carse(2012)]{Carse2012}
James Carse.
\newblock \emph{Finite and infinite games : a vision of life as play and
  possibility}.
\newblock Free Press, New York, 2012.
\newblock ISBN 1476731713.

\bibitem[Ambrose(2021{\natexlab{a}})]{Ambrose2021}
Mitch Ambrose.
\newblock Senators pump brakes on {Endless Frontier Act}, Apr
  2021{\natexlab{a}}.
\newblock URL
  \url{https://www.aip.org/fyi/2021/senators-pump-brakes-endless-frontier-act}.

\bibitem[van Dalen et~al.(2014)van Dalen, Mehmood, Verstraten, and van~der
  Wiel]{Dalen2014}
Ryanne van Dalen, Sultan Mehmood, Pal Verstraten, and Karen van~der Wiel.
\newblock Public funding of science: An international comparison.
\newblock 2014.
\newblock URL
  \url{https://www.cpb.nl/en/publication/public-funding-of-science-an-international-comparison}.

\bibitem[Abbott and Schiermeier(2019)]{Abbott2019}
Alison Abbott and Quirin Schiermeier.
\newblock How {European} scientists will spend {\mbox{\texteuro}}100 billion.
\newblock \emph{Nature}, 569\penalty0 (7757):\penalty0 472--475, May 2019.
\newblock \doi{10.1038/d41586-019-01566-z}.
\newblock URL \url{https://doi.org/10.1038/d41586-019-01566-z}.

\bibitem[Ambrose(2021{\natexlab{b}})]{Ambrose2021may}
Mitch Ambrose.
\newblock Senate taking {Endless Frontier Act} dispute into floor debate, May
  2021{\natexlab{b}}.
\newblock URL
  \url{https://www.aip.org/fyi/2021/senate-taking-endless-frontier-act-dispute-floor-debate}.

\bibitem[Ambrose(2021{\natexlab{c}})]{Ambrose2021jun}
Mitch Ambrose.
\newblock Halftime for {R\&D} push as {Senate} passes {Endless Frontier} bill,
  Jun 2021{\natexlab{c}}.
\newblock URL
  \url{https://www.aip.org/fyi/2021/halftime-rd-push-senate-passes-endless-frontier-bill}.

\end{thebibliography}

\end{document}